\pdfoutput=1
\RequirePackage{ifpdf}
\ifpdf % We are running pdfTeX in pdf mode
\documentclass[pdftex]{sigma}
\else
\documentclass{sigma}
\fi

\numberwithin{equation}{section}

\usepackage{mathtools} % extension of amsmath that also fixes two bugs

\usepackage{enumitem} % e.g. [label=\emph{\roman*}), resume]

\usepackage{bbm} % for \mathbbm{1}
\usepackage{wasysym} % for \eighthnote and \rightturn

\usepackage[dvipsnames]{xcolor}

\usepackage{tikz}
	\usetikzlibrary{decorations.markings} % placing arrows halfway lines
	\usetikzlibrary{patterns} % hatch for wall

\DeclareMathOperator{\E}{e}
\newcommand{\I}{\mathrm{i}}

\DeclarePairedDelimiter{\bra}{\langle}{\rvert}
\DeclarePairedDelimiter{\ket}{\lvert}{\rangle}
\DeclarePairedDelimiterX{\braket}[2]{\langle}{\rangle}{#1\vert#2}

\DeclareMathOperator*{\ordprod}{\prod\limits^{\vbox to -.25ex{\kern-.6ex\hbox{$\leftharpoonup$}\vss}}}
\DeclareMathOperator*{\ordprodopp}{\prod\limits^{\vbox to -.25ex{\kern-.6ex\hbox{$\rightharpoonup$}\vss}}}

\DeclareMathOperator{\Up}{
	\,
	\tikz[baseline={([yshift=-.5*10pt*0.6]current bounding box.center)},scale=0.6,decoration={markings, mark=at position 0.7 with {\arrow[scale=1.25,>=stealth]{>}}}]{ % 0.59375 = 0.5+.75*1.25
	\draw[postaction=decorate] (0,0) -- (0,.6); }
	\,
}
\DeclareMathOperator{\Down}{
	\,
	\tikz[baseline={([yshift=-.5*10pt*0.6]current bounding box.center)},scale=0.6,decoration={markings, mark=at position 0.7 with {\arrow[scale=1.25,>=stealth]{>}}}]{ % 0.59375 = 0.5+.75*1.25
		\draw[postaction=decorate] (0,.6) -- (0,0); }
	\,
}

\DeclareMathOperator{\id}{\mathbbm{1}} %{\mathbf{1}}

\DeclareMathOperator{\sgn}{sgn}

 \let\Im\undefined \DeclareMathOperator{\Im}{Im}

\begin{document}
\allowdisplaybreaks

\newcommand{\arXivNumber}{1801.09635}

\renewcommand{\thefootnote}{}

\renewcommand{\PaperNumber}{064}

\FirstPageHeading
	
\ShortArticleName{The Functional Method for the Domain-Wall Partition Function}
	
\ArticleName{The Functional Method for the Domain-Wall\\ Partition Function\footnote{This paper is a~contribution to the Special Issue on Elliptic Hypergeometric Functions and Their Applications. The full collection is available at \href{https://www.emis.de/journals/SIGMA/EHF2017.html}{https://www.emis.de/journals/SIGMA/EHF2017.html}}}
	
\Author{Jules LAMERS}
	
\AuthorNameForHeading{J.~Lamers}
	
\Address{Department of Mathematical Sciences, Chalmers University of Technology\\ and University of Gothenburg, SE-412 96 G\"oteborg, Sweden}
\Email{\href{mailto:julesl@chalmers.se}{julesl@chalmers.se}}
\URLaddress{\url{http://www.math.chalmers.se/~julesl/home.html}}

\ArticleDates{Received January 30, 2018, in final form June 17, 2018; Published online June 26, 2018}

\Abstract{We review the (algebraic-)functional method devised by Galleas and further developed by Galleas and the author. We first explain the method using the simplest example: the computation of the partition function for the six-vertex model with domain-wall boundary conditions. At the heart of the method lies a linear functional equation for the partition function. After deriving this equation we outline its analysis. The result is a closed expression in the form of a symmetrized sum~-- or, equivalently, multiple-integral formula~-- that can be rewritten to recover Izergin's determinant. Special attention is paid to the relation with other approaches. In particular we show that the Korepin--Izergin approach can be recovered within the functional method. We comment on the functional method's range of applicability, and review how it is adapted to the technically more involved example of the elliptic solid-on-solid model with domain walls and a reflecting end. We present a~new formula for the partition function of the latter, which was expressed as a determinant by Tsuchiya--Filali--Kitanine. Our result takes the form of a `crossing-symmetrized' sum with~$2^L$ terms featuring the elliptic domain-wall partition function, which appears to be new also in the limiting case of the six-vertex model. Further taking the rational limit we recover the expression obtained by Frassek using the boundary perimeter Bethe ansatz.}
	
	
\Classification{82B23; 30D05}

\renewcommand{\thefootnote}{\arabic{footnote}}
\setcounter{footnote}{0}

\section{Introduction}\label{s:intro}

The \emph{$($algebraic-$)$functional method} provides a way for analysing and computing key quantities, such as partition functions, for quantum-integrable models. This approach was first put forward by Galleas in 2010~\cite{Gal_10} and subsequently developed by Galleas and the author \cite{Gal_11,Gal_12,Gal_13b,Gal_13a,Gal_14,Gal_15b,GL_14,Lam_15,Lam_16}.

This review is intended to give an overview of the author's work on the functional method. In Section~\ref{s:constr} we explain the method using the simplest example, namely the exact computation of the domain-wall partition function, where it can be treated rigorously~\cite{Lam_16}. The result is a~symmetrized sum that can be recognized the outcome of Baxter's perimeter Bethe ansatz~\cite{Bax_87} for the case of domain-wall boundaries. We pay special attention to the relation to the approach of Korepin--Izergin~\cite{Ize_87, Kor_82}, which can be recovered within the functional method~\cite{Lam_16}, and that of Bogoliubov, Pronko and Zvonarev~\cite{BPZ_02}. This is discussed in Section~\ref{s:comparison}, most of which is new compared to \cite{Lam_16}.

In Section~\ref{s:variations} we discuss the applicability of the functional method. The largest portion of this section is devoted to illustrating how the method can be applied for the more complex setting of the elliptic solid-on-solid model with a reflecting end and domain walls at the other boundaries~\cite{Lam_15,Lam_16}. In Section~\ref{s:new} we further simplify the result, yielding a novel expression of this partition function as a `crossing-symmetrized' sum involving the elliptic domain-wall partition function. Even in the limiting case of the six-vertex model this relation between partition functions is new to the best of our knowledge.

Proofs and computational details can be found in \cite{Lam_16}.\footnote{Our conventions follow \cite{Lam_15,Lam_16}, except that for aesthetic reasons we have chosen to work with $\sin$ instead of $\sinh$, and changed notation to $x_i \coloneqq \lambda_i/\gamma$, $y_j \coloneqq \mu_j/\gamma$, $z\coloneqq \theta/\gamma$ and $\kappa\coloneqq \zeta/\gamma$.}

\section{The domain-wall partition function}\label{s:setup}

To set the scene we give a recap of the six-vertex model, its algebraic description, the definition of the domain-wall partition function, and the exact computation of the latter by Korepin and Izergin. More details can be found in the references, see especially~\cite{Lam_16}.

\subsection{Set-up: the six-vertex model with domain walls}

\begin{figure}[h]
	\centering
\includegraphics{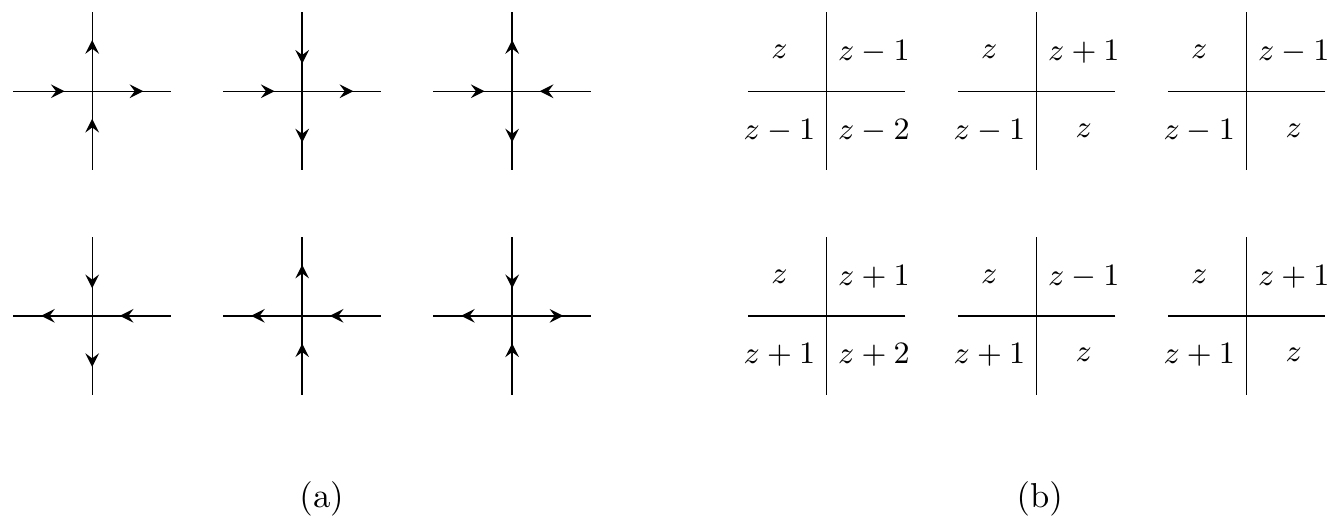}

\caption{(a)~The allowed vertex configurations for the six-vertex model. In the `zero-field' case the corresponding vertex weights, from left to right, are $a$, $b$, $c$, cf.~\eqref{eq:partition_function_general} and \eqref{eq:R-mat}. (b)~The allowed height profiles for Baxter's solid-on-solid model. The local weights are $a_\pm(z)$, $b_\pm(z)$, $c_\pm(z)$, read from left to right and with a `$+$' for the top row, cf.~\eqref{eq:partition_function_general_dynamical}--\eqref{eq:R-mat_dynamical}.}
	\label{fg:six_vertices}
\end{figure}

The six-vertex model is a classical statistical-physical model defined on a square lattice. The microscopic degrees of freedom are arrows pointing in either direction along each edge, subject to the (ice) rule that at every vertex two arrows point in and two point out. This leaves the six allowed arrow configurations around each vertex shown in Fig.~\ref{fg:six_vertices}(a). We focus on the symmetric (`zero-field') case that is invariant under the global reversal of all arrows. The partition function
	\begin{gather}
	Z = \sum_{\substack{\text{arrow}\\\text{configs}}} a^{N_a} b^{N_b}c^{N_c} , \label{eq:partition_function_general}\\
	N_a \coloneqq \#\,\tikz[baseline={([yshift=-.5*12pt*0.5]current bounding box.center)},scale=0.4,decoration={markings, mark=at position 0.65 with {\arrow[scale=.8,>=stealth]{>}}}]{
		\draw[postaction=decorate] (0,1) -- (1,1);
		\draw[postaction=decorate] (1,1) -- (2,1);
		\draw[postaction=decorate] (1,0) -- (1,1);
		\draw[postaction=decorate] (1,1) -- (1,2);
	} \, + \, \#\, \tikz[baseline={([yshift=-.5*12pt*0.5]current bounding box.center)},scale=0.4,decoration={markings, mark=at position 0.65 with {\arrow[scale=.8,>=stealth]{>}}}]{
		\draw[postaction=decorate] (1,1) -- (0,1);
		\draw[postaction=decorate] (2,1) -- (1,1);
		\draw[postaction=decorate] (1,1) -- (1,0);
		\draw[postaction=decorate] (1,2) -- (1,1);
	} , \qquad N_b \coloneqq \#\,\tikz[baseline={([yshift=-.5*12pt*0.5]current bounding box.center)},scale=0.4,decoration={markings, mark=at position 0.65 with {\arrow[scale=.8,>=stealth]{>}}}]{
		\draw[postaction=decorate] (0,1) -- (1,1);
		\draw[postaction=decorate] (1,1) -- (2,1);			
		\draw[postaction=decorate] (1,1) -- (1,0);			
		\draw[postaction=decorate] (1,2) -- (1,1);
	} \, + \, \#\, \tikz[baseline={([yshift=-.5*12pt*0.5]current bounding box.center)},scale=0.4,decoration={markings, mark=at position 0.65 with {\arrow[scale=.8,>=stealth]{>}}}]{
		\draw[postaction=decorate] (1,1) -- (0,1);
		\draw[postaction=decorate] (2,1) -- (1,1);
		\draw[postaction=decorate] (1,0) -- (1,1);
		\draw[postaction=decorate] (1,1) -- (1,2);
	} , \qquad N_c \coloneqq \#\,\tikz[baseline={([yshift=-.5*12pt*0.5]current bounding box.center)},scale=0.4,decoration={markings, mark=at position 0.65 with {\arrow[scale=.8,>=stealth]{>}}}]{
		\draw[postaction=decorate] (0,1) -- (1,1);
		\draw[postaction=decorate] (2,1) -- (1,1);
		\draw[postaction=decorate] (1,1) -- (1,0);
		\draw[postaction=decorate] (1,1) -- (1,2);
	} \, + \, \#\, \tikz[baseline={([yshift=-.5*12pt*0.5]current bounding box.center)},scale=0.4,decoration={markings, mark=at position 0.65 with {\arrow[scale=.8,>=stealth]{>}}}]{
		\draw[postaction=decorate] (1,1) -- (0,1);
		\draw[postaction=decorate] (1,1) -- (2,1);
		\draw[postaction=decorate] (1,0) -- (1,1);
		\draw[postaction=decorate] (1,2) -- (1,1);
	} , \nonumber
	\end{gather}
counts the number of allowed configurations, each with a (Boltzmann) weight that keeps track of the occurring vertices. The result is a polynomial in the vertex weights $a$, $b$, $c$.

Consider a portion of the square lattice consisting of $L$ horizontal lines and $L$ vertical lines. The \emph{inhomogeneous} six-vertex model depends on $L$ spectral parameters~$x_i$ (one for each \emph{horizontal} line of the lattice) and inhomogeneity parameters~$y_j$ (one for each \emph{vertical} line), together with the (`global') crossing parameter~$\gamma \in \mathbf{C}^\times$. We view the $x_i$ as variables and the other parameters as fixed. Abbreviating $[w]\coloneqq \sin(\gamma\,w)$, so that $[w]/[1]$ is the \textit{q}-analogue of $w$ with $q\coloneqq \E^{\I\gamma}$, the three pairs of nonzero vertex weights may be parametrized as
	\begin{gather} \label{eq:vertices}
	a(w) \coloneqq [w+1] , \qquad\qquad b(w) \coloneqq [w] , \qquad\qquad c(w) \coloneqq [1] ,
	\end{gather}
where $w = x_i - y_j$ for the vertex at which the $i$th horizontal line meets the $j$th vertical line. The corresponding partition function depends on all $x_i$, $y_j$ as well as $\gamma$.

{\bf Algebraic formulation.} For each line of the lattice let us introduce a copy of a two-dimen\-sional vector space~$V$ with basis vectors $\ket{\Up}$ and $\ket{\Down}$ labelled by the two directions the arrows on that line can have. Then the weights~\eqref{eq:vertices} for the six vertices in Fig.~\ref{fg:six_vertices}(a) can be encoded in the $R$-matrix
	\begin{gather} \label{eq:R-mat}
	{ \setlength{\arraycolsep}{2pt} % decrease column spacing in arrays (e.g. pmatrix); default: 6pt
	R(w) =
	\begin{pmatrix}	a(w)\! & \color{gray!85}{0} & \color{gray!85}{0} & \color{gray!85}{0} \\ \color{gray!85}{0} & \!b(w) & c(w)\! & \color{gray!85}{0} \\ \color{gray!85}{0} & \!c(w) & b(w)\! & \color{gray!85}{0} \\ \color{gray!85}{0} & \color{gray!85}{0} & \color{gray!85}{0} & \! a(w) \end{pmatrix} ,
	\qquad
	\tikz[baseline={([yshift=-.5*10pt*0.6]current bounding box.center)}, scale=0.6,font=\footnotesize]{
		\draw (0,1) -- (2,1);
		\draw (1,0) -- (1,2);
	}
	 =
	\begin{pmatrix}	
	\tikz[baseline={([yshift=-.5*10pt*0.6]current bounding box.center)},scale=0.3,decoration={markings, mark=at position 0.75 with {\arrow[scale=1,>=stealth]{>}}}]{
		\draw[postaction=decorate] (0,1) -- (1,1);
		\draw[postaction=decorate] (1,1) -- (2,1);
		\draw[postaction=decorate] (1,0) -- (1,1);
		\draw[postaction=decorate] (1,1) -- (1,2);
	}
	& \color{gray!85}{0} & \color{gray!85}{0} & \color{gray!85}{0} \\ \color{gray!85}{0} &
	\tikz[baseline={([yshift=-.5*10pt*0.6]current bounding box.center)},scale=0.3,decoration={markings, mark=at position 0.75 with {\arrow[scale=1,>=stealth]{>}}}]{
		\draw[postaction=decorate] (0,1) -- (1,1);
		\draw[postaction=decorate] (1,1) -- (2,1);			
		\draw[postaction=decorate] (1,1) -- (1,0);			
		\draw[postaction=decorate] (1,2) -- (1,1);
		\draw[white] (1,0) -- (1,-.2);
	}
	&
	\tikz[baseline={([yshift=-.5*10pt*0.6]current bounding box.center)},scale=0.3,decoration={markings, mark=at position 0.75 with {\arrow[scale=1,>=stealth]{>}}}]{
		\draw[postaction=decorate] (1,1) -- (0,1);
		\draw[postaction=decorate] (1,1) -- (2,1);
		\draw[postaction=decorate] (1,0) -- (1,1);
		\draw[postaction=decorate] (1,2) -- (1,1);
		\draw[white] (1,0) -- (1,-.3);
	}
	& \color{gray!85}{0} \\ \color{gray!85}{0} &
	\tikz[baseline={([yshift=-.5*10pt*0.6]current bounding box.center)},scale=0.3,decoration={markings, mark=at position 0.75 with {\arrow[scale=1,>=stealth]{>}}}]{
		\draw[postaction=decorate] (0,1) -- (1,1);
		\draw[postaction=decorate] (2,1) -- (1,1);
		\draw[postaction=decorate] (1,1) -- (1,0);
		\draw[postaction=decorate] (1,1) -- (1,2);
	}
	&
	\tikz[baseline={([yshift=-.5*10pt*0.6]current bounding box.center)},scale=0.3,decoration={markings, mark=at position 0.75 with {\arrow[scale=1,>=stealth]{>}}}]{
		\draw[postaction=decorate] (1,1) -- (0,1);
		\draw[postaction=decorate] (2,1) -- (1,1);
		\draw[postaction=decorate] (1,0) -- (1,1);
		\draw[postaction=decorate] (1,1) -- (1,2);
	}
	& \color{gray!85}{0} \\ \color{gray!85}{0} & \color{gray!85}{0} & \color{gray!85}{0} &
	\tikz[baseline={([yshift=-.5*10pt*0.6]current bounding box.center)},scale=0.3,decoration={markings, mark=at position 0.75 with {\arrow[scale=1,>=stealth]{>}}}]{
		\draw[postaction=decorate] (1,1) -- (0,1);
		\draw[postaction=decorate] (2,1) -- (1,1);
		\draw[postaction=decorate] (1,1) -- (1,0);
		\draw[postaction=decorate] (1,2) -- (1,1);
	}
	\end{pmatrix} .
	}
	\end{gather}
In the diagrammatic version we imagine `time' increasing from the left and bottom to the right and top. The crucial property of the $R$-matrix is that it satisfies the \emph{Yang--Baxter equation},
	\begin{gather} \label{eq:YBE}
	\begin{aligned}
	& R_{ij}(x_i-x_j) R_{ik}(x_i-x_k)R_{jk}(x_j-x_k) \\
& \qquad{}	= R_{jk}(x_j-x_k) R_{ik}(x_i-x_k)R_{ij}(x_i-x_j)
	\end{aligned}
	\qquad \text{or} \qquad \tikz[baseline={([yshift=-.5*10pt*0.6 +8.5pt]current bounding box.center)}, xscale=0.6, yscale=0.5, font=\footnotesize]{
		\pgfmathsetmacro{\csc}{1/sin(130)}
		\draw (-130:1.5*\csc) node[below]{$i$} -- (50:1.5*\csc);
		\draw (.5,-1.5) node[below]{$j$} -- (.5,1.5);
		\draw (-50:1.5*\csc) node[below]{$k$} -- (130:1.5*\csc);
	}
	=
	\tikz[baseline={([yshift=-.5*10pt*0.6 +8.5pt]current bounding box.center)}, xscale=0.6, yscale=0.5, font=\footnotesize]{
		\pgfmathsetmacro{\csc}{1/sin(130)}
		\draw (-130:1.5*\csc) node[below]{$i$} -- (50:1.5*\csc);
		\draw (-.5,-1.5) node[below]{$j$} -- (-.5,1.5);
		\draw (-50:1.5*\csc) node[below]{$k$} -- (130:1.5*\csc);
	} .
	\end{gather}
On the left the subscripts say on which factors of the $2^3$-dimensional vector space $V_i\otimes V_j\otimes V_k$ each $R$-matrix acts nontrivially: $R_{ij} = R\otimes \id$, $R_{jk} = \id \otimes R$, and $R_{ik} = (\id\otimes P)(R\otimes \id)(\id\otimes P) = (P\otimes \id)(\id\otimes R)(P\otimes \id)$ with $P$ the permutation matrix. The diagram is a way to encode the same equation: the `time' specifies the ordering, and the labels at the `incoming' end of each line play the role of the subscripts (whence specifying the arguments, cf.~just below \eqref{eq:vertices}). From the $R$-matrix we construct for each row $i$ the operators
	\begin{gather} \label{eq:ABCD}
	\begin{aligned}
	A(x_i) & \coloneqq
	\tikz[baseline={([yshift=-.5*10pt*0.5+7pt]current bounding box.center)},
	scale=0.5,font=\footnotesize,decoration={markings, mark=at position 0.65 with {\arrow[scale=1.25,>=stealth]{>}}}]{
		\draw[postaction=decorate] (0,1) -- (1,1); \draw (1,1) -- (2.6,1) (3.4,1) -- (4,1); \draw[postaction=decorate] (4,1) -- (5,1);
		\draw (1,0) node[below]{$1$} -- (1,2);
		\draw (2,0) node[below]{$2$} -- (2,2);
		\draw (4,0) node[below]{$L$} -- (4,2);
		\foreach \y in {-1,...,1} \draw (3+.2*\y,1) node{$\cdot\mathstrut$};
		\foreach \y in {-1,...,1} \draw (3+.2*\y,-.4) node{$\cdot\mathstrut$};
	}
 , \qquad
	& B(x_i) \coloneqq
	\tikz[baseline={([yshift=-.5*10pt*0.5+7pt]current bounding box.center)},
	scale=0.5,font=\footnotesize,decoration={markings, mark=at position 0.65 with {\arrow[scale=1.25,>=stealth]{>}}}]{
		\draw[postaction=decorate] (1,1) -- (0,1); \draw (1,1) -- (2.6,1) (3.4,1) -- (4,1); \draw[postaction=decorate] (4,1) -- (5,1);
		\draw (1,0) node[below]{$1$} -- (1,2);
		\draw (2,0) node[below]{$2$} -- (2,2);
		\draw (4,0) node[below]{$L$} -- (4,2);
		\foreach \y in {-1,...,1} \draw (3+.2*\y,1) node{$\cdot\mathstrut$};
		\foreach \y in {-1,...,1} \draw (3+.2*\y,-.4) node{$\cdot\mathstrut$};
	}
 , \\
	C(x_i) & \coloneqq
	\tikz[baseline={([yshift=-.5*10pt*0.5+7pt]current bounding box.center)},
	scale=0.5,font=\footnotesize,decoration={markings, mark=at position 0.65 with {\arrow[scale=1.25,>=stealth]{>}}}]{
		\draw[postaction=decorate](0,1) -- (1,1); \draw (1,1) -- (2.6,1) (3.4,1) -- (4,1); \draw[postaction=decorate] (5,1) -- (4,1);
		\draw (1,0) node[below]{$1$} -- (1,2);
		\draw (2,0) node[below]{$2$} -- (2,2);
		\draw (4,0) node[below]{$L$} -- (4,2);
		\foreach \y in {-1,...,1} \draw (3+.2*\y,1) node{$\cdot\mathstrut$};
		\foreach \y in {-1,...,1} \draw (3+.2*\y,-.4) node{$\cdot\mathstrut$};
	}
 ,
	& D(x_i) \coloneqq
	\tikz[baseline={([yshift=-.5*10pt*0.5+7pt]current bounding box.center)},
	scale=0.5,font=\footnotesize,decoration={markings, mark=at position 0.65 with {\arrow[scale=1.25,>=stealth]{>}}}]{
		\draw[postaction=decorate] (1,1) -- (0,1); \draw (1,1) -- (2.6,1) (3.4,1) -- (4,1); \draw[postaction=decorate] (5,1) -- (4,1);
		\draw (1,0) node[below]{$1$} -- (1,2);
		\draw (2,0) node[below]{$2$} -- (2,2);
		\draw (4,0) node[below]{$L$} -- (4,2);
		\foreach \y in {-1,...,1} \draw (3+.2*\y,1) node{$\cdot\mathstrut$};
		\foreach \y in {-1,...,1} \draw (3+.2*\y,-.4) node{$\cdot\mathstrut$};
	} .
	\end{aligned}
	\end{gather}
Thus $B(x_i) = \bra{
\tikz[baseline={([yshift=-.5*10pt*0.6]current bounding box.center)},scale=0.6,decoration={markings, mark=at position 0.7 with {\arrow[scale=1.25,>=stealth]{>}}}]{ % 0.59375 = 0.5+.75*1.25
		\draw[postaction=decorate] (0,0) -- (.6,0); }
}\, R_{iL}(x_i-y_L) \cdots R_{i1}(x_i-y_1)\, \ket{\tikz[baseline={([yshift=-.5*10pt*0.6]current bounding box.center)},scale=0.6,decoration={markings, mark=at position 0.7 with {\arrow[scale=1.25,>=stealth]{>}}}]{ % 0.59375 = 0.5+.75*1.25
	\draw[postaction=decorate] (.6,0) -- (0,0); }
}$, and so on. These operators act on the $2^L$-dimensional vector space $\bigotimes_{j=1}^L V_j$ associated to the $L$ vertical lines. A simple example of such an action (on a dual vector) is given in \eqref{eq:bra_A} below. Since the $R$-matrix~\eqref{eq:R-mat} satisfies the Yang--Baxter equation \eqref{eq:YBE} the operators~\eqref{eq:ABCD} obey commutation rules that are conveniently encoded in the `$RTT$-relations'
	\begin{gather} \label{eq:RTT}
	\tikz[baseline={([yshift=-.5*10pt*0.5+7pt]current bounding box.center)},% 7 pt = scriptsize using text size 10 pt
	scale=0.5,font=\footnotesize]{
		\draw (1,0) node[below]{$1$} -- (1,4);
		\draw (2,0) node[below]{$2$} -- (2,4);
		\draw (4,0) node[below]{$L$} -- (4,4);
		\foreach \x in {-1,...,1} \foreach \y in {1,2} \draw (3+.2*\x,\y) node{$\cdot\mathstrut$};
		\draw (0,1) node[left]{$i'$} -- (2.6,1); \draw[rounded corners=5pt] (3.4,1) -- (5,1) -- (6,3) -- (6.5,3);
		\draw (0,2) node[left]{$i\hphantom{'}$} -- (2.6,2); \draw (3.4,2) -- (6.5,2);
		\foreach \y in {-1,...,1} \draw (3+.2*\y,-.4) node{$\cdot\mathstrut$};
	}
	 =
	\tikz[baseline={([yshift=-.5*10pt*0.5+7pt]current bounding box.center)}, % 7 pt = scriptsize using text size 10 pt
	scale=0.5, font=\footnotesize]{
		\draw (2.5,-1) node[below]{$1$} -- (2.5,3);
		\draw (3.5,-1) node[below]{$2$} -- (3.5,3);
		\draw (5.5,-1) node[below]{$L$} -- (5.5,3);
		\foreach \x in {-1,...,1} \foreach \y in {1,2} \draw (4.5+.2*\x,\y) node{$\cdot\mathstrut$};
		\draw[rounded corners=5pt] (0,0) node[left]{$i'$} -- (.5,0) -- (1.5,2) -- (3,2) -- (4.1,2) (4.9,2) -- (6.5,2);
		\draw (0,1) node[left]{$i\hphantom{'}$} -- (4.1,1) (4.9,1) -- (6.5,1);
		\foreach \y in {-1,...,1} \draw (4.5+.2*\y,-1.4) node{$\cdot\mathstrut$};
	}.
	\end{gather}
These are the defining relations of the \emph{Yang--Baxter algebra}, with the entries of the $R$-matrix $R_{i,i'}(x_i-x_{i'})$ playing the role of structure constants: each choice of arrows on the four horizontal external edges gives one such relation. (The bends in the horizontal lines have no significance, only their crossings do.) For example, fixing these arrows all outwards as in $\tikz[baseline={([yshift=-.5*10pt*0.6]current bounding box.center)},scale=0.5,decoration={markings, mark=at position 0.7 with {\arrow[scale=.8,>=stealth]{>}}}]{ % 0.59375 = 0.5+.75*1.25
	\draw[postaction=decorate] (.6,0) -- (0,0);
	\draw[postaction=decorate] (.6,.6) -- (0,.6);
	\draw[postaction=decorate] (1.4,0) -- (2,0);
	\draw[postaction=decorate] (1.4,.6) -- (2,.6);
	\foreach \y in {-1,...,1} \draw (1+.2*\y,.3) node{$\cdot\mathstrut$};
}$ gives
	\begin{gather} \label{eq:BB_pre}
	a(x_i-x_{i'}) B(x_i) B(x_{i'}) = a(x_i-x_{i'}) B(x_{i'}) B(x_i) ,
	\end{gather}
while reversing the bottom left arrow as in
$\tikz[baseline={([yshift=-.5*10pt*0.6]current bounding box.center)},scale=0.5,decoration={markings, mark=at position 0.7 with {\arrow[scale=.8,>=stealth]{>}}}]{ % 0.59375 = 0.5+.75*1.25
	\draw[postaction=decorate] (0,0) -- (.6,0);
	\draw[postaction=decorate] (.6,.6) -- (0,.6);
	\draw[postaction=decorate] (1.4,0) -- (2,0);
	\draw[postaction=decorate] (1.4,.6) -- (2,.6);
	\foreach \y in {-1,...,1} \draw (1+.2*\y,.3) node{$\cdot\mathstrut$};
}$ yields
	\begin{gather} \label{eq:AB_pre}
	a(x_i-x_{i'}) B(x_i) A(x_{i'}) = b(x_i-x_{i'}) A(x_{i'}) B(x_i) + c(x_i-x_{i'}) B(x_{i'}) A(x_i) .
	\end{gather}
These are the relations that we will need below.

{\bf Domain walls.} As with any statistical-physical model the goal from the viewpoint of physics is to study the thermodynamics for macroscopically large systems. The usual strategy is to first compute the partition function for finite but arbitrary system size~$L$ and subsequently study its asymptotic behaviour as $L\to\infty$. Along the way one has to choose some boundary conditions. For periodic boundaries, so that the lattice is wrapped around a torus, \eqref{eq:partition_function_general} can be studied using the Bethe ansatz: see, e.g.,~\cite{Lam_14} and the references therein. This converts the computation of the partition function to the problem of solving a system of coupled algebraic equations, the Bethe-ansatz equations. The latter give enough information as $L\to\infty$ to obtain exact expressions for macroscopic quantities such as the bulk free energy~\cite{Lie_67b,Lie_67c,Lie_67a,Sut_67}.

One naively expects the thermodynamic properties to be insensitive to the choice of boundary conditions used at the intermediate step of finite systems. It came as a great surprise that for the six-vertex model the thermodynamics \emph{does} depend on the choice of boundary conditions, as Korepin and Zinn-Justin discovered in 2000~\cite{KZ_00} while studying the partition function for a~particular configuration of fixed arrows on the boundary known as `domain walls':
	\begin{gather} \label{eq:partition_function}
	Z(\vec{x}\,) \coloneqq Z(x_1,\dots,x_L) \coloneqq \bra{{\color{Red}\Down\cdots\Down}} B(x_1) \cdots B(x_L) \ket{{\color{Cerulean}\Up\cdots\Up}} =
	\begin{tikzpicture}[baseline={([yshift=-.5*10pt*0.5]current bounding box.center)}, font=\small, scale=0.5, decoration={markings, mark=at position 0.65 with {\arrow[scale=1.25,>=stealth]{>}}}] % 0.59375 = 0.5+.75*1.25
		\foreach \y in {1,2,4} {
			\draw[postaction=decorate] (1,\y) -- (0,\y);
			\draw (1,\y) -- (2.6,\y) (3.4,\y) -- (4,\y);
			\foreach \x in {-1,...,1} \draw (3+.2*\x,\y) node{$\cdot\mathstrut$};
			\draw[postaction=decorate] (4,\y) -- (5,\y);
		}
		\foreach \x in {1,2,4} {
			\draw[postaction=decorate,color=Cerulean] (\x,0) -- (\x,1);
			\draw (\x,1) -- (\x,2.6) (\x,3.4) -- (\x,4);
			\foreach \y in {-1,...,1} \draw (\x,3+.2*\y) node{$\cdot\mathstrut$};
			\draw[postaction=decorate,color=Red] (\x,5) -- (\x,4);
		}
	\end{tikzpicture} .
	\end{gather}
(Here and later on the colours only serve to highlight the structure of the expressions.) This object, known as the \emph{domain-wall partition function}, was introduced by Korepin 1982~\cite{Kor_82}. Unlike for periodic boundary conditions, which have only been solved exactly in the limit of macroscopically large systems, the domain-wall partition function can be computed exactly for any finite system size~$L$. This is the topic of the next sections.

\subsection{Context: Korepin--Izergin method} \label{s:KI}

Let us write $Z_L$ when we want to emphasize the system size under consideration; sometimes we will also explicitly indicate the dependence on the inhomogeneities~$\vec{y}\coloneqq (y_1,\dots,y_L)$. Korepin~\cite{Kor_82} showed that (for generic values of the inhomogeneities) the domain-wall partition function~\eqref{eq:partition_function} is characterized by \emph{analytic properties},\footnote{These properties are readily obtained from \eqref{eq:partition_function}. Symmetry in the $x_i$ holds by~\eqref{eq:BB_pre}. The polynomial property follows by noting that domain walls require at least one~$c_-$ in every row, so there are at most $L-1$ vertices $a(x_i-y_j)$ and $b(x_i-y_j)$, which by \eqref{eq:vertices} are trigonometric monomials of degree one in~$x_i$.}
\begin{itemize}\itemsep=0pt
	\item $Z$ is doubly symmetric: it is a symmetric function in the $x_i$, and in the parameters $y_j$;
	\item $Z_L$ is a trigonometric polynomial of degree $L-1$ in each variable $x_i$: it equals $\E^{-(L-1)\mspace{1mu}\I\gamma\,x_i}$ times a polynomial of degree $L-1$ in $\E^{2 \mspace{1mu}\I\gamma\, x_i}$ for each $i$,
\end{itemize}
together with a \emph{recurrence relation} between partition functions for successive system sizes,
	\begin{gather} \label{eq:Korepin}
	Z_L(\vec{x};\vec{y}\,)\big|_{x_1 = y_1} = \text{factor} \times Z_{L-1}(x_2,\dots,x_L;y_2,\dots,y_L) ,
	\end{gather}
with initial condition $Z_1(x) =
\tikz[baseline={([yshift=-.5*10pt*0.6]current bounding box.center)},scale=0.3,decoration={markings, mark=at position 0.75 with {\arrow[scale=1.25,>=stealth]{>}}}]{
	\draw[postaction=decorate] (1,1) -- (0,1);
	\draw[postaction=decorate] (1,1) -- (2,1);
	\draw[postaction=decorate,color=Cerulean] (1,0) -- (1,1);
	\draw[postaction=decorate,color=Red] (1,2) -- (1,1);
}$. The factor in \eqref{eq:Korepin} is a monomial in the vertex weights depending on all parameters of the model; since its explicit form is not relevant for us here we defer its explicit form, and the simple proof of \eqref{eq:Korepin}, to Section~\ref{s:comparison}. Since \eqref{eq:Korepin}, together with its analogues for $x_i=y_j$ (by double symmetry), fixes $Z_L$ at sufficiently many distinct points (for generic values of the $y_j$) these properties uniquely determine $Z_L$ for each $L$ by Lagrange interpolation, see again Section~\ref{s:comparison}.

Five years later Izergin~\cite{Ize_87,ICK_92} came up with a beautiful answer involving an $L\times L$ determinant:
	\begin{gather} \label{eq:Izergin}
	Z_L(\vec{x};\vec{y}\,) = [1]^L \frac{\prod\limits_{i,j=1}^L [x_i-y_j+1 ,x_i-y_j]}{\prod\limits_{i<j}^L [x_i-x_j ,y_j-y_i]} \times \det_{i,j} \left( \frac{1}{[x_i-y_j+1 , x_i-y_j]} \right) ,
	\end{gather}
where we use the shorthand
	\begin{gather} \label{eq:shorthand}
	[w_1 ,w_2,\dots] \coloneqq [w_1] [w_2]\cdots .
	\end{gather}

It is not hard to check that the function on the right-hand side of \eqref{eq:Izergin} meets all requirements, which proves that it is indeed the domain-wall partition function. In the homogeneous limit, where the $x_i - y_j \to w$ become independent of $i$ and $j$, the (Hankel) determinant obtained from \eqref{eq:Izergin} can be used to compute the bulk free energy as $L\to\infty$ \cite{KZ_00,Zin_00}.

\section{Functional method} \label{s:constr}

Next we turn to the functional method, which consists of two steps:
\begin{enumerate}[label=\arabic*)]\itemsep=0pt
	\item use the algebraic structure underlying the model to derive a functional equation satisfied by the partition function;
	\item analyse this functional equation to find a recipe for obtaining a closed formula for the partition function.
\end{enumerate}
The first step is quite straightforward, the second step more involved -- rather more so than the simple analysis involved in the Korepin--Izergin method from Section~\ref{s:KI}. As we will see, however, the functional approach in fact contains the method of Korepin--Izergin.

Before we commence let us briefly consider the \emph{cyclic} functional equation
	\begin{gather} \label{eq:cyclic}
	\sum_{j=1}^n F(z_j,z_{j+1},\dots,z_{j+n-1}) = 0
	\end{gather}
over $\mathbb{C}^n$, where we identify $z_{j+n} \equiv z_j$ for all $j$. It is not hard to see that the solution must be of the form $F(z_1,\dots,z_n) = G(z_1,\dots,z_n) - G(z_2,\dots,z_n,z_1)$ for some function~$G$. Reversely, for any choice of $G$ this combination obeys \eqref{eq:cyclic}. We see that a single (linear) functional equation may admit many linearly independent solutions also in reasonable function spaces.

\subsection{From algebra to functional equations} \label{s:derivation}

The starting point for deriving the functional equation is the algebraic expression~\eqref{eq:partition_function} for the domain-wall partition function.

Because of the ice rule the operator $A(x_0)$ from \eqref{eq:ABCD} preserves the numbers of up- and down-pointing arrows. In particular, $\bra{{\color{Red}\Down\cdots\Down}}$ is a (dual) eigenvector:
	\begin{gather}
	\bra{{\color{Red}\Down\cdots\Down}} A(x_0) =
	\tikz[baseline={([yshift=-.5*10pt*0.5]current bounding box.center)},
	scale=0.5,font=\scriptsize,decoration={markings, mark=at position 0.65 with {\arrow[scale=1.25,>=stealth]{>}}}]{
		\draw (1,0) -- (1,1); \draw[postaction=decorate,color=Red] (1,2) -- (1,1);
		\draw (2,0) -- (2,1); \draw[postaction=decorate,color=Red] (2,2) -- (2,1);
		\draw (4,0) -- (4,1); \draw[postaction=decorate,color=Red] (4,2) -- (4,1);
		\foreach \y in {-1,...,1} \draw (3+.2*\y,1) node{$\cdot\mathstrut$};
		\draw[postaction=decorate] (0,1) -- (1,1); \draw (1,1) -- (2.6,1) (3.4,1) -- (4,1); \draw[postaction=decorate] (4,1) -- (5,1);
	}
	 = \bra{{\color{Red}\Down}}\otimes \tikz[baseline={([yshift=-.5*10pt*0.5]current bounding box.center)},
	scale=0.5,font=\scriptsize,decoration={markings, mark=at position 0.65 with {\arrow[scale=1.25,>=stealth]{>}}}]{
		\draw[postaction=decorate,color=Red] (1,1) -- (1,0); \draw[postaction=decorate] (1,2) -- (1,1);
		\draw (2,0) -- (2,1); \draw[postaction=decorate] (2,2) -- (2,1);
		\draw (4,0) -- (4,1); \draw[postaction=decorate] (4,2) -- (4,1);
		\foreach \y in {-1,...,1} \draw (3+.2*\y,1) node{$\cdot\mathstrut$};
		\draw[postaction=decorate] (0,1) -- (1,1); \draw[postaction=decorate] (1,1) -- (2,1); \draw (2,1) -- (2.6,1) (3.4,1) -- (4,1); \draw[postaction=decorate] (4,1) -- (5,1);
	}
	= \cdots \nonumber\\
\hphantom{\bra{{\color{Red}\Down\cdots\Down}} A(x_0)}{} =
	\underbrace{
	\tikz[baseline={([yshift=-.5*10pt*0.5]current bounding box.center)},
	scale=0.5,font=\scriptsize,decoration={markings, mark=at position 0.65 with {\arrow[scale=1.25,>=stealth]{>}}}]{
		\draw[postaction=decorate,color=Red] (1,1) -- (1,0); \draw[postaction=decorate] (1,2) -- (1,1);
		\draw[postaction=decorate,color=Red] (2,1) -- (2,0); \draw[postaction=decorate] (2,2) -- (2,1);
		\draw[postaction=decorate,color=Red] (4,1) -- (4,0); \draw[postaction=decorate] (4,2) -- (4,1);
		\draw[postaction=decorate] (0,1) -- (1,1); \draw[postaction=decorate] (1,1) -- (2,1); \draw[postaction=decorate] (2,1) -- (3,1); \draw[postaction=decorate] (3,1) -- (4,1); \draw[postaction=decorate] (4,1) -- (5,1);
		\fill[white] (2.6,.5) rectangle (3.4,1.5);
		\foreach \y in {-1,...,1} \draw (3+.2*\y,1) node{$\cdot\mathstrut$};
	}
	}_{\textcolor{Red}{\text{eigenvalue}_{\mspace{2mu}0}}}
	\times \bra{{\color{Red}\Down\cdots\Down}} .\label{eq:bra_A}
	\end{gather}
Therefore we may insert $A(x_0)$ on the left of the $B$s in \eqref{eq:partition_function} at the cost of its \textcolor{Red}{\text{eigenvalue}$_{\mspace{2mu}0}$}:
	\begin{gather} \label{eq:derivation_1}
	\textcolor{Red}{\text{eigenvalue}_{\mspace{2mu}0}} \times Z(\vec{x}\,) = \bra{{\color{Red}\Down\cdots\Down}} A(x_0) B(x_1) \cdots B(x_L) \ket{{\color{Cerulean}\Up\cdots\Up}} .
	\end{gather}

Now we can move the $A$ past all $B$s by rewriting \eqref{eq:AB_pre} to get a relation of the form
	\begin{gather} \label{eq:AB}
	A(x) B(x') = \substack{\text{some}\\\text{structure}\\\text{constant}} \times B(x') A(x) + \substack{\text{another} \\ \text{structure}\\\text{constant}} \times B(x) A(x') .
	\end{gather}
In the first term on the right-hand side the $A$ just moves past the $B$~-- up to a structure constant coming from the $R$-matrices in~\eqref{eq:RTT}~-- while in the second term $A$ exchanges spectral parameter with the $B$ it passes~-- again up to a structure constant. Thus the result of moving the~$A$ all the way to the right of the $B$s in~\eqref{eq:derivation_1} is a linear combination of terms for which the $A$ ends up with any of the spectral parameters,
	\begin{gather} \label{eq:derivation_2}
	\textcolor{Red}{\text{eigenvalue}_{\mspace{2mu}0}} \times Z(\vec{x}\,) = \sum_{\nu=0}^L \text{coefficient}_\nu \times
	\bra{{\color{Red}\Down\cdots\Down}} \prod_{\substack{\rho=0 \\ \hphantom{\rho}\neq \nu}}^L B(x_\rho) A(x_\nu) \ket{{\color{Cerulean}\Up\cdots\Up}} .
	\end{gather}
The $L+1$ coefficients come from the structure constants in \eqref{eq:AB} and are easy to find when one exploits the fact that $Z$ is symmetric in the $x_i$: this computation is routine in the context of the algebraic Bethe ansatz.

Finally we use that $\ket{{\color{Cerulean}\Up\cdots\Up}}$ is an eigenvector of $A$ too~-- the computation resembles \eqref{eq:bra_A}~-- so~$A(x_\nu)$ may be replaced by its \textcolor{Cerulean}{\text{eigenvalue}$_\nu$}. This allows us to recognize the domain-wall partition function~\eqref{eq:partition_function}, depending on $L$ out of the $L+1$ $x$s, in each term in~\eqref{eq:derivation_2} to get
	\begin{gather} \label{eq:derivation_3}
	\textcolor{Red}{\text{eigenvalue}_{\mspace{2mu}0}} \times Z(\vec{x}\,) = \sum_{\nu=0}^L \textcolor{Cerulean}{\text{eigenvalue}_\nu} \times \text{coefficient}_\nu \times Z(x_0,\dots,\widehat{x_\nu},\dots,x_L) ,
	\end{gather}
where the caret indicates that $x_\nu$ is omitted.

{\bf Result.} The domain-wall partition function obeys the linear functional equation~\cite{Gal_13b}
	\begin{gather} \label{eq:functional}
	\sum_{\nu=0}^L M_\nu(x_0;\vec{x}\,) Z(x_0,\dots,\widehat{x_\nu},\dots,x_L) = 0 ,
	\end{gather}
where the coefficients obtained from~\eqref{eq:derivation_3} explicitly read
\begin{subequations} \label{eq:M_0i}
	\begin{align}
	M_0(x_0;\vec{x}\,) & \coloneqq \!\!\!\!\!\!\!\! \underbrace{\prod_{j=1}^L \! b(x_0-y_j)}_\text{\textcolor{Red}{eigenvalue$_{\mspace{2mu}0}$} for $\bra{{\color{Red}\Down\cdots\Down}}$\qquad } \!\!\!\!\! - \! \underbrace{\prod_{j=1}^L \! a(x_0-y_j)}_\text{\textcolor{Cerulean}{eigenvalue$_{\mspace{2mu}0}$} for $\ket{{\color{Cerulean}\Up\cdots\Up}}$} \!\! \underbrace{\prod_{j=1}^L \! \frac{a(x_j-x_0)}{b(x_j-x_0)}}_\text{coefficient$_{\mspace{2mu}0}$} , \label{eq:M_0} \\
	M_i(x_0;\vec{x}\,) & \coloneqq \!\!\!\!\!\! \underbrace{\prod_{j=1}^L \! a(x_i-y_j)}_\text{\textcolor{Cerulean}{eigenvalue$_i$} for $\ket{{\color{Cerulean}\Up\cdots\Up}}$\quad } \!\!\!\! \underbrace{\frac{c(x_i-x_0)}{b(x_i-x_0)} \prod_{\substack{j=1 \\ \hphantom{j}\neq i}}^L \! \frac{a(x_j-x_i)}{b(x_j-x_i)}}_\text{coefficient$_i$} , \qquad 1\leq i\leq L . \label{eq:M_i}
	\end{align}
\end{subequations}
This concludes the first step from the beginning of Section~\ref{s:constr}.

\subsection{Analysis and solution}\label{s:analysis}

We reserve the notation `$Z$' for the domain-wall partition function~\eqref{eq:partition_function} and study the functional equation
	\begin{gather} \label{eq:functional_2}
	\sum_{\nu=0}^L M_\nu(x_0;\vec{x}\,) F(x_0,\dots,\widehat{x_\nu},\dots,x_L) = 0 ,
	\end{gather}
for an unknown function $F$ that we wish to find. The coefficients are \eqref{eq:M_0i}, i.e.,
\begin{subequations} \label{eq:M_0i_alt}
	\begin{gather}
	M_0(x_0;\vec{x}\,) = \prod_{j=1}^L [x_0-y_j] - \prod_{j=1}^L [x_0-y_j+1] \frac{[x_j-x_0+1]}{[x_j-x_0]} , \label{eq:M_0_alt} \\
	M_i(x_0;\vec{x}\,) = \frac{[1]}{[x_i-x_0]} \prod_{j=1}^L [x_i-y_j+1] \prod_{\substack{j=1 \\ \hphantom{j}\neq i}}^L \! \frac{[x_j-x_i+1]}{[x_j-x_i]} , \qquad 1\leq i\leq L . \label{eq:M_i_alt}
	\end{gather}
\end{subequations}
Our goals are to understand whether one may forget about the origin of \eqref{eq:functional_2} in order to analyse it, to which extent Korepin's characterization from Section~\ref{s:KI} can be recovered from it, and of course whether it can be solved. Let us present the highlights of this analysis.

A first nice property of~\eqref{eq:functional_2} is that sufficiently well-behaved solutions to the equation necessarily share analytic properties with the domain-wall partition function. Indeed, it can be shown that if $F$ is meromorphic then it is symmetric in the $x_i$. Furthermore, if it is a trigonometric polynomial in $x_i$, which already follows from \eqref{eq:partition_function_general}--\eqref{eq:vertices} but is also reasonable from \eqref{eq:M_0i_alt}, then it has degree $L-1$. By symmetry of the coefficients \eqref{eq:M_0i_alt} in the $y_j$, permutations of the inhomogeneities map solutions to solutions; so if the solution turns out to be unique then it must be symmetric in the $y_j$. Thus we do not require the analytic properties from Section~\ref{s:KI} as input, but can seek solutions of the~\eqref{eq:functional_2} in a rather general space of functions.

Since the number of variables in \eqref{eq:functional_2} is larger than the number of arguments of $F$ we can specialize any single variable to a convenient value.

By \eqref{eq:M_0i_alt} the greatest simplification occurs when $x_0 = x_\star = y_k - 1$ for some $1\leq k\leq L$. For $L=1$ this specialization implies that $F=F_1$ is a constant, independent of $x_1$. The value of this constant is not determined by the functional equation, which after all is linear in $F$.

In general under this specialization \eqref{eq:functional_2} can be solved for $F(\vec{x}\,)$ to give
	\begin{gather}
	F(\vec{x}\,) = \sum_{j=1}^L\frac{M_j(x_\star;\vec{x}\,)}{M_0(x_\star;\vec{x}\,)} \times F(\vec{x})\big|_{x_j = x_\star} \nonumber\\
\hphantom{F(\vec{x}\,)}{} = \sum_{j=1}^L\frac{M_j(y_k-1;\vec{x}\,)}{M_0(y_k-1;\vec{x}\,)} \prod_{\substack{i=1\\ \smash{\hphantom{i}\neq j}}}^L [x_i-y_k] \times \tilde{F}(x_1,\dots,\widehat{x_j},\dots,x_L) , \label{eq:towards_recipe}
	\end{gather}
where the first equality uses the symmetry of $F$ and the second equality is a consequence of the so-called `special zeroes' that any $F$ satisfying \eqref{eq:functional_2} possesses,\footnote{Namely:\label{footnote} if $F$ obeys \eqref{eq:functional_2} then it vanishes whenever any two of its variables are set to $y_k-1$ and $y_k$ respectively. On the right-hand side of~\eqref{eq:towards_recipe} we have extracted a factor from $F$ making this property manifest. For the domain-wall partition function it is a direct consequence of Korepin's recurrence relation: $y_k-1$ is a zero of the factor in~\eqref{eq:Korepin}, see~\eqref{eq:Korepin_derivation}. However, it may also be proven solely from~\eqref{eq:functional_2} using the symmetry of~$F$~\cite{Lam_16}.} yielding a factor times a~$\tilde{F}$ is a~trigonometric polynomial of degree $L-2$ in each of its $L-1$ arguments.

When we set $x_L = y_k$ in \eqref{eq:towards_recipe} only the $L$th term in the second line survives and we precisely obtain Korepin's recurrence relation~\eqref{eq:Korepin} if we could interpret $\tilde{F}$ as the solution for the functional equation for $Z_{L-1}$. To see whether such an interpretation is allowed we plug~\eqref{eq:recipe} into~\eqref{eq:functional_2} to find that if~$F$ obeys \eqref{eq:functional_2} then $\tilde{F}$ does indeed obey the functional equation for $Z_{L-1}$ with inhomogeneities~$y_1,\dots,\widehat{y_k},\dots,y_L$. In other words, solutions to~\eqref{eq:functional} for successive~$L$ are recursively related by Korepin's recurrence relation~\eqref{eq:Korepin}!

We thus recover all ingredients from the Korepin--Izergin-method (except for the initial condition) within our approach, based solely on \eqref{eq:functional} for a quite general space of functions in which we look for solutions.

Moreover, \eqref{eq:towards_recipe} now gives a \emph{recipe} that allows us to construct $F_L=F$ in terms of $F_{L-1}=\tilde{F}$:
\begin{gather}
	F_L(\vec{x};\vec{y}\,) = \text{const} \times \sum_{j=1}^L M_j(y_k-1;\vec{x}\,) \prod_{\substack{i=1\\ \smash{\hphantom{i}\neq j}}}^L [x_i-y_k]\nonumber\\
\hphantom{F_L(\vec{x};\vec{y}\,)=}
 \times F_{L-1}(x_1,\dots,\widehat{x_j},\dots,x_L;y_1,\dots,\widehat{y_k},\dots,y_L) \nonumber \\
\hphantom{F_L(\vec{x};\vec{y}\,)}	 = \text{const} \times [1] \sum_{j=1}^L \prod_{\substack{i=1 \\ \hphantom{i}\neq k}}^L [x_j-y_i+1] \prod_{\substack{i=1 \\ \hphantom{i}\neq j}}^L [x_i-y_k] \frac{[x_i-x_j+1]}{[x_i-x_j]} \nonumber \\
\hphantom{F_L(\vec{x};\vec{y}\,)=} \times F_{L-1}(x_1,\dots,\widehat{x_j},\dots,x_L;y_1,\dots,\widehat{y_k},\dots,y_L) , \label{eq:recipe}
\end{gather}
where we absorbed the (constant) denominators $M_0(y_n-1;x_{\sigma 1},\dots,x_{\sigma n}\,)$ in the prefactor. Unlike~\eqref{eq:Korepin} this recipe does not involve any specialization on the left-hand side. In particular, the recipe expresses any $F_L$ in terms of some $F_{L-1}$. Recall that the solution for $L=1$ is just a constant. Thus the recipe uniquely determines $F_2$ up to a constant normalization factor. Repeating this argument it follows that the functional equation~\eqref{eq:functional_2} has, up to normalization, a~\emph{unique} solution in the space of trigonometric polynomials in $L$ variables. This is a nontrivial property for a functional equation, as the example of the cyclic functional equation~\eqref{eq:cyclic} shows.

By iterating \eqref{eq:recipe} we readily find a closed formula for $F_L$. The iteration will require a~different choice of~$k$ at each step, exhausting all inhomogeneities along the way. By the solution's uniqueness up to normalization different choices will at most affect the constant prefactor. For later reference we give the result obtained by iteration using the first expression on the right-hand side of~\eqref{eq:recipe}:
	\begin{gather} \label{eq:F_via_mn}
	F_L(\vec{x};\vec{y}\,) = \Omega_L \sum_{\sigma\in S_L} \prod_{n=1}^L m_n(y_n -1;x_{\sigma 1},\dots,x_{\sigma n}) \prod_{i<j}^L [x_{\sigma i} - y_j ] ,
	\end{gather}
where $\Omega_L$ is an normalization constant and $m_n$ denotes the coefficient $M_L$ from \eqref{eq:M_i_alt} at `length'~$L=n$. A simple way to carry out this iteration is to compute the term with $\sigma=e$ in~\eqref{eq:F_via_mn} by taking $k=n$ and picking up the last term ($j=n$) from the sum (so that the products in the last expression in~\eqref{eq:recipe} have the same range), at each application of~\eqref{eq:recipe}; then~\eqref{eq:F_via_mn} follows by symmetry in the $x_i$.

Fixing $\Omega_L=1$ in \eqref{eq:F_via_mn} in order to match the value of the domain-wall partition function at any single point\footnote{The correct overall normalization can be found from the leading behaviour as the spectral parameters tend to infinity~\cite{Gal_10,Gal_11,Gal_12,Gal_13b,Gal_13a,Gal_14,Gal_15b,GL_14}, or more simply by evaluation at a particular point, e.g., $x_i=y_i$ for all $i$~\cite{Lam_16}.}~-- this is our boundary condition~-- we thus obtain a formula in terms of a~symmetrized sum using the functional method~\cite{Gal_11}:
	\begin{gather} \label{eq:symmetrized_sum}
	Z_L(\vec{x};\vec{y}\,) = [1]^L \sum_{\sigma\in S_L} \prod_{i<j}^L [x_{\sigma i}-y_j , x_{\sigma j}-y_i +1] \frac{[x_{\sigma i}-x_{\sigma j}+1]}{[x_{\sigma i}-x_{\sigma j}]} .
	\end{gather}
Although there are $L!$ terms, one for each permutation $\sigma\in S_L$, this is still much better than the $2^{(L-1)^2}$ terms that the domain-wall partition function naively contains, cf.~\eqref{eq:partition_function_general}, \eqref{eq:partition_function}. To see that the poles at coinciding spectral parameters are removable rewrite \eqref{eq:symmetrized_sum} as
	\begin{gather} \label{eq:symmetrized_sum_vandermonde}
	Z_L(\vec{x};\vec{y}\,) = \frac{[1]^L}{\prod\limits_{i<j}^L [x_i-x_j]} \sum_{\sigma\in S_L} \sgn(\sigma) \prod_{i<j}^L [x_{\sigma i}-y_j , x_{\sigma j}-y_i +1 ,x_{\sigma i}-x_{\sigma j}+1] .
	\end{gather}
Since the sum is explicitly \emph{anti}symmetrized it is divisible by the Vandermonde factor in the denominator. The symmetry in the $x_i$, obvious from the symmetrization in \eqref{eq:symmetrized_sum}, makes it possible~\cite{Lam_15,Lam_16} to rewrite the result in terms of a repeated contour integral to obtain a~multiple-integral formula. Note that the symmetry in the inhomogeneities~$y_j$ is not manifest in \smash{\eqref{eq:symmetrized_sum}--\eqref{eq:symmetrized_sum_vandermonde}}.

One can recognize \eqref{eq:symmetrized_sum} as the outcome of Baxter's `perimeter Bethe ansatz'~\cite{Bax_87} for the special case of a square lattice with domain-wall boundaries. Although \eqref{eq:symmetrized_sum} is not obviously equal to Izergin's determinant~\eqref{eq:Izergin}, one can directly show that the two expressions coincide using a version of Lagrange interpolation~\cite{Lam_16}. (The latter is also the way in which Rosengren~\cite{Ros_09} first found his sum of determinants for the partition function of the elliptic solid-on-solid model with domain-wall boundary conditions starting from a symmetrized sum.)

\subsection{Relation with other approaches} \label{s:comparison}

The preceding analysis tells us that
	\begin{gather} \label{eq:functional_implies_KI}
	\begin{cases}
	\text{functional equation~\eqref{eq:functional_2}}, \\
	\text{space of trig polynomials}, \\
	\text{boundary condition (for each $L$)}
	\end{cases}
	\!\!\!\!\!\! \implies
	\begin{cases}
	\text{Korepin's recurrence relation~\eqref{eq:Korepin}}, \\
	\text{doubly symmetric}, \\
	\text{trig polynomial, degree $<L$ in each $x_i$}, \\
	\text{initial condition ($L=1$)}.
	\end{cases}\!\!\!\!\!\!\!\!
	\end{gather}
Actually, the functional method gives us more: \eqref{eq:recipe} gives a recipe to obtain \eqref{eq:F_via_mn} by straightforward recursion. Instead, Korepin's characterization from Section~\ref{s:KI} features a specialization on the right-hand side of \eqref{eq:Korepin}. Before we show how one can proceed in this approach let us give the simple derivation of Korepin's recurrence relation based on \eqref{eq:partition_function}:
	\begin{gather}
	Z_L(\vec{x};\vec{y}\,)\big|_{x_1=y_1} =
	\begin{tikzpicture}[baseline={([yshift=-.5*10pt*0.5]current bounding box.center)}, font=\small, scale=0.5, decoration={markings, mark=at position 0.65 with {\arrow[scale=1.25,>=stealth]{>}}}] % 0.59375 = 0.5+.75*1.25
		\foreach \y in {1,2,4,5} {
			\draw[postaction=decorate] (1,\y) -- (0,\y);
			\draw (1,\y) -- (2.6,\y) (3.4,\y) -- (5,\y);
			\foreach \x in {-1,...,1} \draw (3+.2*\x,\y) node{$\cdot\mathstrut$};
			\draw[postaction=decorate] (5,\y) -- (6,\y);
		}
		\foreach \x in {1,2,4,5} {
			\draw[postaction=decorate,color=Cerulean] (\x,0) -- (\x,1);
			\draw (\x,1) -- (\x,2.6) (\x,3.4) -- (\x,5);
			\foreach \y in {-1,...,1} \draw (\x,3+.2*\y) node{$\cdot\mathstrut$};
			\draw[postaction=decorate,color=Red] (\x,6) -- (\x,5);
		}
		\draw[postaction=decorate] (1,1) -- (2,1);
		\draw[postaction=decorate] (1,2) -- (1,1);
	\end{tikzpicture}
	 =
	\begin{tikzpicture}[baseline={([yshift=-.5*10pt*0.5]current bounding box.center)}, font=\small, scale=0.5, decoration={markings, mark=at position 0.65 with {\arrow[scale=1.25,>=stealth]{>}}}] % 0.59375 = 0.5+.75*1.25
		\draw[postaction=decorate] (1,2) -- (1,1);
		\foreach \x in {2,4,5} \draw[postaction=decorate,color=Cerulean] (\x,1) -- (\x,2);
		\foreach \x in {1,...,4} \draw[postaction=decorate] (\x,1) -- (\x+1,1);
		\foreach \y in {2,4,5} \draw[postaction=decorate] (2,\y) -- (1,\y);
		\foreach \y in {1,...,4} \draw[postaction=decorate] (1,\y+1) -- (1,\y);
		\draw[white,fill=white] (.5,2.6) rectangle (1.5,3.4) (2.6,.5) rectangle (3.4,1.5);
		\foreach \y in {1,2,4,5} {
			\draw[postaction=decorate] (1,\y) -- (0,\y);
			\draw (2,\y) -- (2.6,\y) (3.4,\y) -- (5,\y);
			\foreach \x in {-1,...,1} \draw (3+.2*\x,\y) node{$\cdot\mathstrut$};
			\draw[postaction=decorate] (5,\y) -- (6,\y);
		}
		\foreach \x in {1,2,4,5} {
			\draw[postaction=decorate,color=Cerulean] (\x,0) -- (\x,1);
			\draw (\x,2) -- (\x,2.6) (\x,3.4) -- (\x,5);
			\foreach \y in {-1,...,1} \draw (\x,3+.2*\y) node{$\cdot\mathstrut$};
			\draw[postaction=decorate,color=Red] (\x,6) -- (\x,5);
		}
	\end{tikzpicture}
	 =
	\begin{tikzpicture}[baseline={([yshift=-.5*10pt*0.5]current bounding box.center)}, font=\small, scale=0.5, decoration={markings, mark=at position 0.65 with {\arrow[scale=1.25,>=stealth]{>}}}] % 0.59375 = 0.5+.75*1.25
		\begin{scope}[shift={(2.5,2.5)}]
			\foreach \y in {1,3,4} {
				\draw[postaction=decorate] (1,\y) -- (0,\y);
				\draw (1,\y) -- (1.6,\y) (2.4,\y) -- (4,\y);
				\foreach \x in {-1,...,1} \draw (2+.2*\x,\y) node{$\cdot\mathstrut$};
				\draw[postaction=decorate] (4,\y) -- (5,\y);
			}
			\foreach \x in {1,3,4} {
				\draw[postaction=decorate,color=Cerulean] (\x,0) -- (\x,1);
				\draw (\x,1) -- (\x,1.6) (\x,2.4) -- (\x,4);
				\foreach \y in {-1,...,1} \draw (\x,2+.2*\y) node{$\cdot\mathstrut$};
				\draw[postaction=decorate,color=Red] (\x,5) -- (\x,4);
			}
		\end{scope}
		\begin{scope}[xshift=2.5cm]
			\foreach \x in {1,3,4} {
				\draw[postaction=decorate,color=Cerulean] (\x,0) -- (\x,1);
				\draw[postaction=decorate,color=Cerulean] (\x,1) -- (\x,2);
			}
			\foreach \x in {0,...,4} \draw[postaction=decorate] (\x,1) -- (\x+1,1);
			\draw[white,fill=white] (1.6,.5) rectangle (2.4,1.5);
			\foreach \x in {-1,...,1} \draw (2+.2*\x,1) node{$\cdot\mathstrut$};
		\end{scope}
		\draw[postaction=decorate] (1,1) -- (0,1);
		\draw[postaction=decorate,color=Cerulean] (1,0) -- (1,1);
		\draw[postaction=decorate] (1,1) -- (2,1);
		\draw[postaction=decorate] (1,2) -- (1,1);
		\begin{scope}[yshift=2.5cm]
			\foreach \y in {1,3,4} {
				\draw[postaction=decorate] (1,\y) -- (0,\y);
				\draw[postaction=decorate] (2,\y) -- (1,\y);
			}
			\foreach \y in {0,...,4} \draw[postaction=decorate] (1,\y+1) -- (1,\y);
			\draw[white,fill=white] (.5,1.6) rectangle (1.5,2.4);
			\foreach \y in {-1,...,1} \draw (1,2+.2*\y) node{$\cdot\mathstrut$};
		\end{scope}
	\end{tikzpicture} \nonumber\\
\hphantom{Z_L(\vec{x};\vec{y}\,)\big|_{x_1=y_1} }{} = c(0) \prod_{i=2}^L a(x_i-y_1) a(y_1-y_i) \times Z_{L-1}(x_2,\dots,x_L;y_2,\dots,y_L) .
\label{eq:Korepin_derivation}
	\end{gather}
The first equality follows since setting $x_L=y_L$ forces the upper right vertex to be a $c_-$: the boundary conditions only allow for $b_+$ or $c_-$ in that corner, and $b(0)=0$ by~\eqref{eq:vertices}. Due to the ice rule this freezes the arrows on the upper row and right-most column, cf.~\eqref{eq:bra_A}. The weight of the $2L-1$ frozen vertices the right-hand side is the factor that we suppressed in~\eqref{eq:Korepin}.

These recurrence relations, together with the analytic properties and initial condition, can be solved by Lagrange interpolation as follows. A basis for the $L$-dimensional space of trigonometric polynomials in $x$ of degree at most $L-1$ is given by $\phi_j(x) \coloneqq \prod\limits_{i(\neq j)}^L [x-y_i]$, which are linearly independent since only $\phi_j$ is nonzero at $x=y_j$. The Lagrange interpolation formula expresses~$Z_L$ in terms of this basis. Focussing on the dependence on $x_1$ it reads
\begin{gather}
	 Z_L(\vec{x};\vec{y}\,) = \sum_{j=1}^L Z_L(\vec{x};\vec{y}\,)\big|_{x_1=y_j} \frac{\phi_j(x_1)}{\phi_j(y_j)} = [1] \sum_{j=1}^L Z_{L-1}(x_2,\dots,x_L;y_1,\dots,\widehat{y_j},\dots,y_L) \nonumber\\
\hphantom{Z_L(\vec{x};\vec{y}\,) =}{} \times \prod_{i=2}^L [x_i -y_j+1] \prod_{\substack{i=1 \\ \hphantom{i}\neq j}}^L [y_j - y_i +1] \frac{[x_1-y_i]}{[y_j-y_i]} ,\label{eq:Lagrange_interpolation}
\end{gather}
where in the second equality we plugged in \eqref{eq:Korepin_derivation} and its analogues through symmetry in the~$y_j$. Repeating \eqref{eq:Lagrange_interpolation} for $x_2,\dots,x_{L-1}$ and using the initial condition gives another symmetrized sum:
	\begin{gather} \label{eq:Lagrange_result}
	Z_L(\vec{x};\vec{y}\,) = [1]^L \sum_{\sigma\in S_L} \prod_{i<j}^L [x_i-y_{\sigma j} ,x_j-y_{\sigma i}+1] \frac{[y_{\sigma i} - y_{\sigma j} +1]}{[y_{\sigma i}-y_{\sigma j}]} .
	\end{gather}
This time the symmetry is manifest for the inhomgeneities but not for the spectral parameters.

If we set $x_i = y'_i -1$ and $y_j = x'_j$ (or better: $y_j = x'_j -\pi/\gamma$, as we will see momentarily) in~\eqref{eq:Lagrange_result} and drop the primes we get back to~\eqref{eq:symmetrized_sum}. The domain-wall partition function does indeed satisfy
	\begin{gather} \label{eq:x_y_symmetry}
	Z(\vec{y}-1;\vec{x}\,) = Z(\vec{y}-1;\vec{x}-\pi/\gamma) = Z(\vec{x};\vec{y}\,) ,
	\end{gather}
as is evident in Izergin's formula~\eqref{eq:Izergin}. To understand this symmetry from the model's set-up recall that we made a~choice in Section~\ref{s:setup}: we sliced up the lattice from \eqref{eq:partition_function} in rows, cf.~\eqref{eq:ABCD}. Of course we may equally well slice it up into columns instead; then the $x_i$ play the role of inhomogeneities while the $y_j$ become our variables. We get back to the other choice by rotating all pictures over $90^\circ$ and reversing the direction of all arrows. Fig.~\ref{fg:six_vertices} shows that this operation amounts to swapping the vertices of weight $a$ and $b$. But at the level of the parametrization~\eqref{eq:vertices} this swap can be implemented by $w \mapsto \pi/\gamma -(w+1)$, which is precisely the effect of setting $x_i= y'_i -1$, $y_j = x'_j -\pi/\gamma$ on $w=x_i - y_j$. Now, Korepin's characterization from Section~\ref{s:KI} also allows for Lagrange interpolation in the $y_j$, with respect to which $Z$ is also a trigonometric polynomial of degree at most $L-1$. The corresponding Lagrange interpolation, using \eqref{eq:Korepin_derivation} with $x_1$ instead of~$y_1$ on the right-hand side, gives{\samepage
\begin{gather}
	 Z_L(\vec{x};\vec{y}\,) = \sum_{j=1}^L Z_L(\vec{x};\vec{y}\,)\big|_{y_1=x_j} \prod_{\substack{i=1 \\ \hphantom{i}\neq j}}^L \frac{[y_1-x_i]}{[x_j-x_i]} = [1] \sum_{j=1}^L Z_{L-1}(x_1,\dots,\widehat{x_j},\dots,x_L;y_2,\dots,y_L) \nonumber\\
\hphantom{Z_L(\vec{x};\vec{y}\,) =}{} \times \prod_{i=2}^L [x_j - y_i +1] \prod_{\substack{i=1 \\ \hphantom{i}\neq j}}^L [x_i -x_j+1] \frac{[y_1-x_i]}{[x_j-x_i]} , \label{eq:Lagrange_interpolation_y}
\end{gather}
should be compared with \eqref{eq:recipe} for $k=1$ and (thus) yields~\eqref{eq:symmetrized_sum}.}

The property \eqref{eq:x_y_symmetry} moreover explains the presence of a second set of Korepin recurrence relations, which may be found as in~\eqref{eq:Korepin_derivation}:
	\begin{gather}
	Z_L(\vec{x};\vec{y}\,)\big|_{x_L = y_1-1} =
	\begin{tikzpicture}[baseline={([yshift=-.5*10pt*0.5]current bounding box.center)}, font=\small, scale=0.5, decoration={markings, mark=at position 0.65 with {\arrow[scale=1.25,>=stealth]{>}}}] % 0.59375 = 0.5+.75*1.25
	\begin{scope}[rotate=180]
	\foreach \y in {1,2,4,5} {
		\draw[postaction=decorate] (1,\y) -- (0,\y);
		\draw (1,\y) -- (2.6,\y) (3.4,\y) -- (5,\y);
		\foreach \x in {-1,...,1} \draw (3+.2*\x,\y) node{$\cdot\mathstrut$};
		\draw[postaction=decorate] (5,\y) -- (6,\y);
	}
	\foreach \x in {1,2,4,5} {
		\draw[postaction=decorate,color=Red] (\x,0) -- (\x,1);
		\draw (\x,1) -- (\x,2.6) (\x,3.4) -- (\x,5);
		\foreach \y in {-1,...,1} \draw (\x,3+.2*\y) node{$\cdot\mathstrut$};
		\draw[postaction=decorate,color=Cerulean] (\x,6) -- (\x,5);
	}
	\draw[postaction=decorate] (5,2) -- (5,1);
	\draw[postaction=decorate] (5,1) -- (4,1);
	\end{scope}
	\end{tikzpicture}
	 =
	\begin{tikzpicture}[baseline={([yshift=-.5*10pt*0.5]current bounding box.center)}, font=\small, scale=0.5, decoration={markings, mark=at position 0.65 with {\arrow[scale=1.25,>=stealth]{>}}}] % 0.59375 = 0.5+.75*1.25
	\begin{scope}[rotate=180]
	\foreach \x in {1,2,4} \draw[postaction=decorate,color=Red] (\x,1) -- (\x,2);
	\foreach \x in {1,...,4} \draw[postaction=decorate] (\x+1,1) -- (\x,1);
	\foreach \y in {2,4,5} \draw[postaction=decorate] (4,\y) -- (5,\y);
	\foreach \y in {1,...,4} \draw[postaction=decorate] (5,\y+1) -- (5,\y);
	\draw[white,fill=white] (4.5,2.6) rectangle (5.5,3.4) (2.6,.5) rectangle (3.4,1.5);
	\foreach \y in {1,2,4,5} {
		\draw[postaction=decorate] (1,\y) -- (0,\y);
		\draw (1,\y) -- (2.6,\y) (3.4,\y) -- (4,\y);
		\foreach \x in {-1,...,1} \draw (3+.2*\x,\y) node{$\cdot\mathstrut$};
		\draw[postaction=decorate] (5,\y) -- (6,\y);
	}
	\foreach \x in {1,2,4,5} {
		\draw[postaction=decorate,color=Red] (\x,0) -- (\x,1);
		\draw (\x,2) -- (\x,2.6) (\x,3.4) -- (\x,5);
		\foreach \y in {-1,...,1} \draw (\x,3+.2*\y) node{$\cdot\mathstrut$};
		\draw[postaction=decorate,color=Cerulean] (\x,6) -- (\x,5);
	}
	\end{scope}
	\end{tikzpicture}
	 =
	\begin{tikzpicture}[baseline={([yshift=-.5*10pt*0.5]current bounding box.center)}, font=\small, scale=0.5, decoration={markings, mark=at position 0.65 with {\arrow[scale=1.25,>=stealth]{>}}}] % 0.59375 = 0.5+.75*1.25
	\begin{scope}[rotate=180]
	\begin{scope}[yshift=2.5cm]
		\foreach \y in {1,3,4} {
			\draw[postaction=decorate] (1,\y) -- (0,\y);
			\draw (1,\y) -- (2.6,\y) (3.4,\y) -- (4,\y);
			\foreach \x in {-1,...,1} \draw (3+.2*\x,\y) node{$\cdot\mathstrut$};
			\draw[postaction=decorate] (4,\y) -- (5,\y);
		}
		\foreach \x in {1,2,4} {
			\draw[postaction=decorate,color=Red] (\x,0) -- (\x,1);
			\draw (\x,1) -- (\x,1.6) (\x,2.4) -- (\x,4);
			\foreach \y in {-1,...,1} \draw (\x,2+.2*\y) node{$\cdot\mathstrut$};
			\draw[postaction=decorate,color=Cerulean] (\x,5) -- (\x,4);
		}
	\end{scope}
	\foreach \x in {1,2,4} \draw[postaction=decorate,color=Red] (\x,0) -- (\x,1);
	\foreach \x in {0,...,4} \draw[postaction=decorate] (\x+1,1) -- (\x,1);
	\draw[white,fill=white] (2.6,.5) rectangle (3.4,1.5);
	\foreach \x in {-1,...,1} \draw (3+.2*\x,1) node{$\cdot\mathstrut$};
	\foreach \x in {1,2,4} \draw[postaction=decorate,color=Red] (\x,1) -- (\x,2);
	\begin{scope}[xshift=1.5cm]
		\draw[postaction=decorate] (5,2) -- (5,1);
		\draw[postaction=decorate,color=Red] (5,0) -- (5,1);
		\draw[postaction=decorate] (5,1) -- (4,1);
		\draw[postaction=decorate] (5,1) -- (6,1);
	\end{scope}
	\begin{scope}[shift={(1.5,2.5)}]
		\foreach \y in {1,3,4} \draw[postaction=decorate] (4,\y) -- (5,\y);
		\foreach \y in {0,...,4} \draw[postaction=decorate] (5,\y+1) -- (5,\y);
		\draw[white,fill=white] (4.5,1.6) rectangle (5.5,2.4);
		\foreach \y in {1,3,4} \draw[postaction=decorate] (5,\y) -- (6,\y);
		\foreach \y in {-1,...,1} \draw (5,2+.2*\y) node{$\cdot\mathstrut$};
		\draw[postaction=decorate,color=Cerulean] (5,5) -- (5,4);
	\end{scope}
	\end{scope}
	\end{tikzpicture} \nonumber\\
\hphantom{Z_L(\vec{x};\vec{y}\,)\big|_{x_L = y_1-1}}{} = c(-1) \prod_{i=1}^{L-1} b(x_i-y_1) \prod_{i=2}^L b(y_1-1-y_i)\nonumber\\
\hphantom{Z_L(\vec{x};\vec{y}\,)\big|_{x_L = y_1-1}=}{} \times Z_{L-1}(x_1,\dots,x_{L-1};y_2,\dots,y_L). \label{eq:Korepin_alt}
	\end{gather}
This relation will be obtained within the functional method by setting $x_i=y'_i-1$, $y_j=x_j'$ in~\eqref{eq:functional_2} and taking $y'_0=y'_\star = x'_L$ and later $y'_1=x'_L+1$. One can again solve~\eqref{eq:Korepin_derivation} by Lagrange interpolation, starting with~$x_L$ and using the basis $\varphi_j(w)\coloneqq \prod\limits_{i(\neq j)}^L [w-y_j+1]$ to get~\eqref{eq:Lagrange_result}.

The functional equation thus somehow `knows' about Lagrange interpolation. A variant of the Korepin--Izergin method for which the same is true was found in~\cite{BPZ_02, KMT_99}. Let us outline the derivation of~\cite{BPZ_02}. The ice rule implies that in the lattice of the domain-wall partition function~\eqref{eq:partition_function} out of the horizontal edges that are adjacent to the left boundary exactly \emph{one} has an arrow pointing to the right. Therefore
\begin{subequations}
	\begin{gather} \label{eq:BPZ}
	 Z_L(\vec{x};\vec{y}\,) =
	\sum_{i=1}^L
	\begin{tikzpicture}[baseline={([yshift=-.5*10pt*0.5]current bounding box.center)}, font=\small, scale=0.5, decoration={markings, mark=at position 0.65 with {\arrow[scale=1.25,>=stealth]{>}}}] % 0.59375 = 0.5+.75*1.25
	\begin{scope}[rotate=180]
	\foreach \y in {1,2,4,5} {
		\draw[postaction=decorate] (1,\y) -- (0,\y);
		\draw (1,\y) -- (2.6,\y) (3.4,\y) -- (5,\y);
		\foreach \x in {-1,...,1} \draw (3+.2*\x,\y) node{$\cdot\mathstrut$};
		\draw[postaction=decorate] (5,\y) -- (6,\y);
	}	
	\foreach \x in {1,2,4,5} {
		\draw[postaction=decorate,color=Red] (\x,0) -- (\x,1);
		\draw (\x,1) -- (\x,2.6) (\x,3.4) -- (\x,5);
		\foreach \y in {-1,...,1} \draw (\x,3+.2*\y) node{$\cdot\mathstrut$};
		\draw[postaction=decorate,color=Cerulean] (\x,6) -- (\x,5);
	}
	\draw[postaction=decorate] (5,2) -- (4,2);
	\node at (6.35,2) {$i$};
	\end{scope}
	\end{tikzpicture}
	 =
	\sum_{i=1}^L
	\begin{tikzpicture}[baseline={([yshift=-.5*10pt*0.5]current bounding box.center)}, font=\small, scale=0.5, decoration={markings, mark=at position 0.65 with {\arrow[scale=1.25,>=stealth]{>}}}] % 0.59375 = 0.5+.75*1.25
	\begin{scope}[rotate=180]
	\foreach \y in {1,4,5} \draw[postaction=decorate] (4,\y) -- (5,\y);
	\foreach \y in {2,...,4} \draw[postaction=decorate] (5,\y+1) -- (5,\y);
	\draw[postaction=decorate] (5,1) -- (5,2);
	\draw[white,fill=white] (4.5,2.6) rectangle (5.5,3.4) (2.6,.5) rectangle (3.4,1.5);
	\foreach \y in {1,2,4,5} {
		\draw[postaction=decorate] (1,\y) -- (0,\y);
		\draw (1,\y) -- (2.6,\y) (3.4,\y) -- (4,\y);
		\foreach \x in {-1,...,1} \draw (3+.2*\x,\y) node{$\cdot\mathstrut$};
		\draw[postaction=decorate] (5,\y) -- (6,\y);
	}	
	\foreach \x in {1,2,4,5} {
		\draw[postaction=decorate,color=Red] (\x,0) -- (\x,1);
		\draw (\x,1) -- (\x,2.6) (\x,3.4) -- (\x,5);
		\foreach \y in {-1,...,1} \draw (\x,3+.2*\y) node{$\cdot\mathstrut$};
		\draw[postaction=decorate,color=Cerulean] (\x,6) -- (\x,5);
	}
	\draw[postaction=decorate] (5,2) -- (4,2);
	\node at (6.35,2) {$i$};
	\end{scope}
	\end{tikzpicture}
	 .
\end{gather}
Each \looseness=1 term on the right is given by a factor, which is read off like in \eqref{eq:Korepin_derivation} and \eqref{eq:Korepin_alt}, times $\bra{{\color{Red}\Down\cdots\Down}} \prod\limits_{j(>i)}^L B_{L-1}(x_j) A_{L-1}(x_i) \prod\limits_{j(<i)} B_{L-1}(x_j) \ket{{\color{Cerulean}\Up\cdots\Up}}$ with operators \eqref{eq:ABCD} for length $L-1$. Setting $x_1=y_1$ in \eqref{eq:BPZ} kills all terms with $i\neq 1$ and the equation boils down to \eqref{eq:Korepin_derivation}. Alternatively one can proceed like in Section~\ref{s:derivation}, using the commutation rule~\eqref{eq:AB} to move the $A$ in each term all the way to the side to exchange it for its eigenvalue, yielding a relation of the form
\begin{gather}
	 Z_L(\vec{x};\vec{y}\,) = \sum_{i=1}^L \text{coefficient}_i \times Z_{L-1}(x_1,\dots,\widehat{x_i},\dots,x_L;y_2,\dots,y_L) .
	\end{gather}
\end{subequations}	
In fact, \looseness=1 one can check that the relation obtained in this way is precisely \eqref{eq:Lagrange_interpolation_y}~\cite{BPZ_02}. For each~$i$ one can of course equally well choose to move the $A$ in~\eqref{eq:BPZ} all the way to the left. Our functional equation~\eqref{eq:functional_2} is the consistency condition ensuring the equality of the results for either choice.

{\samepage To conclude this discussion we note that another Korepin recurrence relation is easily obtained in a~similar fashion to~\eqref{eq:Korepin_derivation} and~\eqref{eq:Korepin_alt}:
	\begin{gather} \label{eq:Korepin_derivation_equiv}
	Z_L(\vec{x};\vec{y}\,)\big|_{x_L=y_L} =
	\begin{tikzpicture}[baseline={([yshift=-.5*10pt*0.5]current bounding box.center)}, font=\small, scale=0.5, decoration={markings, mark=at position 0.65 with {\arrow[scale=1.25,>=stealth]{>}}}] % 0.59375 = 0.5+.75*1.25
	\foreach \y in {1,2,4,5} {
		\draw[postaction=decorate] (1,\y) -- (0,\y);
		\draw (1,\y) -- (2.6,\y) (3.4,\y) -- (5,\y);
		\foreach \x in {-1,...,1} \draw (3+.2*\x,\y) node{$\cdot\mathstrut$};
		\draw[postaction=decorate] (5,\y) -- (6,\y);
	}
	\foreach \x in {1,2,4,5} {
		\draw[postaction=decorate,color=Cerulean] (\x,0) -- (\x,1);
		\draw (\x,1) -- (\x,2.6) (\x,3.4) -- (\x,5);
		\foreach \y in {-1,...,1} \draw (\x,3+.2*\y) node{$\cdot\mathstrut$};
		\draw[postaction=decorate,color=Red] (\x,6) -- (\x,5);
	}
	\draw[postaction=decorate] (5,4) -- (5,5);
	\draw[postaction=decorate] (5,5) -- (4,5);
	\end{tikzpicture}
	 =
	\begin{tikzpicture}[baseline={([yshift=-.5*10pt*0.5]current bounding box.center)}, font=\small, scale=0.5, decoration={markings, mark=at position 0.65 with {\arrow[scale=1.25,>=stealth]{>}}}] % 0.59375 = 0.5+.75*1.25
	\draw[postaction=decorate] (5,4) -- (5,5);
	\foreach \x in {1,2,4} \draw[postaction=decorate,color=Red] (\x,5) -- (\x,4);
	\foreach \x in {1,...,4} \draw[postaction=decorate] (\x+1,5) -- (\x,5);
	\foreach \y in {1,2,4} \draw[postaction=decorate] (4,\y) -- (5,\y);
	\foreach \y in {1,...,4} \draw[postaction=decorate] (5,\y) -- (5,\y+1);
	\draw[white,fill=white] (4.5,2.6) rectangle (5.5,3.4) (2.6,4.5) rectangle (3.4,5.5);
	\foreach \y in {1,2,4,5} {
		\draw[postaction=decorate] (1,\y) -- (0,\y);
		\draw (1,\y) -- (2.6,\y) (3.4,\y) -- (4,\y);
		\foreach \x in {-1,...,1} \draw (3+.2*\x,\y) node{$\cdot\mathstrut$};
		\draw[postaction=decorate] (5,\y) -- (6,\y);
	}
	\foreach \x in {1,2,4,5} {
		\draw[postaction=decorate,color=Cerulean] (\x,0) -- (\x,1);
		\draw (\x,1) -- (\x,2.6) (\x,3.4) -- (\x,4);
		\foreach \y in {-1,...,1} \draw (\x,3+.2*\y) node{$\cdot\mathstrut$};
		\draw[postaction=decorate,color=Red] (\x,6) -- (\x,5);
	}
	\end{tikzpicture}
	 =
	\begin{tikzpicture}[baseline={([yshift=-.5*10pt*0.5]current bounding box.center)}, font=\small, scale=0.5, decoration={markings, mark=at position 0.65 with {\arrow[scale=1.25,>=stealth]{>}}}] % 0.59375 = 0.5+.75*1.25
	\foreach \y in {1,2,4} {
		\draw[postaction=decorate] (1,\y) -- (0,\y);
		\draw (1,\y) -- (2.6,\y) (3.4,\y) -- (4,\y);
		\foreach \x in {-1,...,1} \draw (3+.2*\x,\y) node{$\cdot\mathstrut$};
		\draw[postaction=decorate] (4,\y) -- (5,\y);
	}
	\foreach \x in {1,2,4} {
		\draw[postaction=decorate,color=Cerulean] (\x,0) -- (\x,1);
		\draw (\x,1) -- (\x,2.6) (\x,3.4) -- (\x,4);
		\foreach \y in {-1,...,1} \draw (\x,3+.2*\y) node{$\cdot\mathstrut$};
		\draw[postaction=decorate,color=Red] (\x,5) -- (\x,4);
	}
	\begin{scope}[yshift=1.5cm]
	\foreach \x in {1,2,4} \draw[postaction=decorate,color=Red] (\x,5) -- (\x,4);
	\foreach \x in {0,...,4} \draw[postaction=decorate] (\x+1,5) -- (\x,5);
	\draw[white,fill=white] (2.6,4.5) rectangle (3.4,5.5);
	\foreach \x in {-1,...,1} \draw (3+.2*\x,5) node{$\cdot\mathstrut$};
	\foreach \x in {1,2,4} \draw[postaction=decorate,color=Red] (\x,6) -- (\x,5);
	\end{scope}
	\begin{scope}[shift={(1.5,1.5)}]
	\draw[postaction=decorate] (5,4) -- (5,5);
	\draw[postaction=decorate,color=Red] (5,6) -- (5,5);
	\draw[postaction=decorate] (5,5) -- (4,5);
	\draw[postaction=decorate] (5,5) -- (6,5);
	\end{scope}
	\begin{scope}[xshift=1.5cm]
	\foreach \y in {1,2,4} \draw[postaction=decorate] (4,\y) -- (5,\y);
	\foreach \y in {1,...,4} \draw[postaction=decorate] (5,\y) -- (5,\y+1);
	\draw[white,fill=white] (4.5,2.6) rectangle (5.5,3.4);
	\foreach \y in {1,2,4} \draw[postaction=decorate] (5,\y) -- (6,\y);
	\foreach \y in {-1,...,1} \draw (5,3+.2*\y) node{$\cdot\mathstrut$};
	\draw[postaction=decorate,color=Cerulean] (5,0) -- (5,1);
	\end{scope}
	\end{tikzpicture}.\!\!\!\!\!
	\end{gather}
This recurrence relation is related to a functional equation of `type \textit{D}', derived like in Section~\ref{s:derivation} by inserting $D$ from \eqref{eq:ABCD} rather than $A$ in \eqref{eq:derivation_1}. We did not need this functional equation in Section~\ref{s:analysis}, which can now be understood from the fact that~\eqref{eq:Korepin_derivation_equiv} is equivalent to \eqref{eq:Korepin_derivation} by double symmetry. Another way to see this is that~\eqref{eq:Korepin_derivation_equiv} differs from \eqref{eq:Korepin_derivation} by rotation over $180^\circ$, or applying \eqref{eq:x_y_symmetry} twice; but this does not affect the partition function, which only depends on differences of spectral and inhomogeneity parameters.}

\section{Further examples}\label{s:variations}

\subsection{Comments on applicability}\label{s:applicability}

Having seen how the functional method works in a simple example, and that it is closely related to the Korepin--Izergin method in that case, one may wonder if it can be applied in other settings too: what is the range of applicability of the functional approach?

{\bf Grocery list.} In the preceding section we used the following ingredients to apply the functional method successfully:
\begin{enumerate}[label=\roman*),ref=\roman*]\itemsep=0pt
	\item \label{it:alg_expr} An \emph{algebraic expression} for the quantity of interest. In principle this is available for observables of many models using the solution of the `quantum inverse-scattering problem'~\cite{GK_00, KMT_00,MT_00}.
	\item \label{it:suitable_op} A \emph{suitable operator to insert} in this algebraic expression in such a way that the resulting quantity can be computed in two ways; in particular all terms that arise in the computation should have an `appropriate' form. More concretely, in Section~\ref{s:derivation} this amounted to the existence of an operator for which
	\begin{enumerate}[label=\alph*),ref=\alph*]\itemsep=0pt
		\item both $\bra{{\color{Red}\Down\cdots\Down}}$ and $\ket{{\color{Cerulean}\Up\cdots\Up}}$ are eigenvectors; \label{it:op_eigenvector}
		\item the right-hand side of the relation~\eqref{eq:AB} does not involve new operators that we cannot get rid of. We will get back to this point momentarily. \label{it:suitable_op_b}
	\end{enumerate}
	\item \label{it:comm_symm} The \emph{commutativity} of the $B$s, and reversely, the \emph{symmetry} of reasonable solutions to the functional equation. Perhaps it is possible to relax this condition; here it was convenient and also used to establish the technical but important property that
	\item \label{it:special_zeroes} Reasonable solutions have \emph{`special zeroes'}.\textsuperscript{\,\scriptsize \#\ref{footnote} (p.~\pageref{footnote})}
	\item \label{it:reduction} The `reduced' function $\tilde{F}$ solves the functional equation for $F_{L-1}$, providing the \emph{reduction step} to lower $L$, so that we obtain the recurrence relation from the functional equation.
\end{enumerate}

Note that the Korepin--Izergin method also starts from~\eqref{it:alg_expr} together with some analytic pro\-perties like the symmetry in \eqref{it:comm_symm} that can be surmised from \eqref{it:alg_expr}. Given a Korepin-type recurrence relation it is easy to locate the values of the special zeroes~\eqref{it:special_zeroes}. On the other hand this recurrence relation follows from the functional equation; it is not known whether the existence of special zeroes and a Korepin-type recurrence relation, cf.~\eqref{it:reduction}, are equivalent. Ingredient~\eqref{it:suitable_op} is closely related to the computations from the algebraic Bethe ansatz.

{\bf Other examples.} Luckily, the preceding ingredients are present in several other settings as well. At the end of Section~\ref{s:comparison} we already mentioned that $D$ from~\eqref{eq:ABCD}, which meets require\-ment~\eqref{it:suitable_op} too, can be used instead of $A$ to redo Section~\ref{s:derivation} and obtain another linear functional equation for the domain-wall partition function. (It is also possible to insert $C$; as its relations with $B$ are less simple than~\eqref{eq:AB} the resulting functional equation for $Z$ is rather more complicated~\cite{Gal_10,Gal_11}.)

Another example of ingredient~\eqref{it:suitable_op} is $t = A + D$ in the context of \emph{scalar products} of (\mbox{off-/}on-shell) Bethe vectors, cf.~\cite{Sla_89}, via the functional method~\cite{Gal_14,Gal_15b}. On the other hand, for the domain-wall partition function of the Izergin--Korepin nineteen-vertex model, which was computed exactly at a particular root of unity~\cite{Gar_16}, ingredient \mbox{(\ref{it:suitable_op}\textsuperscript{\,\ref{it:suitable_op_b}})} seems to fail because the analogue of \eqref{eq:AB} mixes the two creation operators $B_1$ and $B_2$ for that model.

Other variations of the setting from Section~\ref{s:setup} to which the functional method can be applied are obtained by
\begin{itemize}\itemsep=0pt
	\item upgrading the six-vertex model to Baxter's solid-on-solid (\textsc{sos}) model~\cite{Gal_12}, and if so,
	\item further refining the \textsc{sos} model to the elliptic case~\cite{Gal_13a,Gal_16b,Gal_16u};
	\item including a reflecting end, with domain walls on the three other boundaries~\cite{GL_14}.
\end{itemize}
The combination of these three options was treated in \cite{Lam_15,Lam_16}: we refer to the resulting quantity as the \emph{reflecting-end partition function} for the elliptic \textsc{sos} model. This is one of the technically most involved examples of the functional method, and our next topic. See \cite{Lam_16} for details.

\subsection{The elliptic reflecting-end partition function} \label{s:SOS}

Baxter's \emph{solid-on-solid} (or \emph{interaction-round-a-face}, \textsc{irf}) model is a generalization of the six-vertex model that is naturally viewed as a height model: the microscopic degrees of freedom are (discrete) height variables associated to the \emph{faces} (plaquettes) of the square lattice, so a~configuration describes a height profile. The heights take values in~$\mathbb{Z}$. Neighbouring heights must differ by one, allowing for the six height profiles around a~vertex shown in Fig.~\ref{fg:six_vertices}(b). There is a~nice correspondence between such height profiles and arrow configurations on the edges \emph{provided} we know the height at any single face of the lattice. The rule to go back and forth between the two settings is the following: going anti-clockwise around a vertex, place an arrow pointing outwards (inwards) if the height increases (decreases) by one. Completing a circle we get back to the height we started at, so the arrows around the vertex must satisfy the ice rule from the six-vertex model. The result is sometimes called a \emph{generalized $($six-$)$vertex model}: it is just a~version of the six-vertex model where we also keep track of (any single, whence all) heights. The partition function is a refinement of~\eqref{eq:partition_function_general}:
	\begin{gather} \label{eq:partition_function_general_dynamical}
	Z = \sum_{\substack{\text{arrow}\\\text{configs}}} \prod_{z \in \mathbb{Z}} a_+(z)^{\#\!\!\tikz[baseline={([yshift=-.5*12pt*0.7]current bounding box.center)},scale=0.25,decoration={markings, mark=at position 0.75 with {\arrow[scale=.7,>=stealth]{>}}},font=\scriptsize]{
		\draw[postaction=decorate] (0,1) -- (1,1);
		\draw[postaction=decorate] (1,1) -- (2,1);
		\draw[postaction=decorate] (1,0) -- (1,1);
		\draw[postaction=decorate] (1,1) -- (1,2);
		\node at (.3,1.7) {$z$};
	}} \, a_-(z)^{\#\!\!\tikz[baseline={([yshift=-.5*12pt*0.7]current bounding box.center)},scale=0.25,decoration={markings, mark=at position 0.75 with {\arrow[scale=.7,>=stealth]{>}}},font=\scriptsize]{
		\draw[postaction=decorate] (1,1) -- (0,1);
		\draw[postaction=decorate] (2,1) -- (1,1);
		\draw[postaction=decorate] (1,1) -- (1,0);
		\draw[postaction=decorate] (1,2) -- (1,1);
		\node at (.3,1.7) {$z$};
	}} \, b_+(z)^{\#\!\!\tikz[baseline={([yshift=-.5*12pt*0.7]current bounding box.center)},scale=0.25,decoration={markings, mark=at position 0.75 with {\arrow[scale=.7,>=stealth]{>}}},font=\scriptsize]{
		\draw[postaction=decorate] (0,1) -- (1,1);
		\draw[postaction=decorate] (1,1) -- (2,1);			
		\draw[postaction=decorate] (1,1) -- (1,0);			
		\draw[postaction=decorate] (1,2) -- (1,1);
		\node at (.3,1.7) {$z$};
	}} \, b_-(z)^{\#\!\!\tikz[baseline={([yshift=-.5*12pt*0.7]current bounding box.center)},scale=0.25,decoration={markings, mark=at position 0.75 with {\arrow[scale=.7,>=stealth]{>}}},font=\scriptsize]{
		\draw[postaction=decorate] (1,1) -- (0,1);
		\draw[postaction=decorate] (2,1) -- (1,1);
		\draw[postaction=decorate] (1,0) -- (1,1);
		\draw[postaction=decorate] (1,1) -- (1,2);
		\node at (.3,1.7) {$z$};
	}} \, c_+(z)^{\#\!\!\tikz[baseline={([yshift=-.5*12pt*0.7]current bounding box.center)},scale=0.25,decoration={markings, mark=at position 0.75 with {\arrow[scale=.7,>=stealth]{>}}},font=\scriptsize]{
		\draw[postaction=decorate] (0,1) -- (1,1);
		\draw[postaction=decorate] (2,1) -- (1,1);
		\draw[postaction=decorate] (1,1) -- (1,0);
		\draw[postaction=decorate] (1,1) -- (1,2);
		\node at (.3,1.7) {$z$};
	}} \, c_-(z)^{\#\!\!\tikz[baseline={([yshift=-.5*12pt*0.7]current bounding box.center)},scale=0.25,decoration={markings, mark=at position 0.75 with {\arrow[scale=.7,>=stealth]{>}}},font=\scriptsize]{
		\draw[postaction=decorate] (1,1) -- (0,1);
		\draw[postaction=decorate] (1,1) -- (2,1);
		\draw[postaction=decorate] (1,0) -- (1,1);
		\draw[postaction=decorate] (1,2) -- (1,1);
		\node at (.3,1.7) {$z$};
	}} .
	\end{gather}

{\bf Algebraic formulation.} We repeat the trick of passing to the inhomogeneous setting with variables~$x_i$ associated to the horizontal lines and parameters~$y_j$ for the vertical lines. The algebraic reformulation is again based on an $R$-matrix containing the weights for the profiles from Fig.~\ref{fg:six_vertices}(b), where we prefer the viewpoint of the generalized vertex model to stress the similarity with~\eqref{eq:R-mat}:
	\begin{gather} \setlength{\arraycolsep}{2pt} % decrease column spacing in arrays (e.g. pmatrix); default: 6pt
	R(w,z) =
	\begin{pmatrix}	a_+(w,z) & \color{gray!85}{0} & \color{gray!85}{0} & \color{gray!85}{0} \\ \color{gray!85}{0} & b_+(w,z) & c_-(w,z) & \color{gray!85}{0} \\ \color{gray!85}{0} & c_+(w,z) & b_-(w,z) & \color{gray!85}{0} \\ \color{gray!85}{0} & \color{gray!85}{0} & \color{gray!85}{0} & a_-(w,z) \end{pmatrix} ,
\nonumber\\
	\tikz[baseline={([yshift=-.5*10pt*0.6]current bounding box.center)}, scale=0.6,font=\footnotesize]{
		\draw[->] (0,1) -- (2,1);
		\draw[->] (1,0) -- (1,2);
		\node at (.5,1.5) {$z$};
	}
	 \coloneqq
	\begin{pmatrix}	
	\tikz[baseline={([yshift=-.5*10pt*0.6]current bounding box.center)},scale=0.3,decoration={markings, mark=at position 0.75 with {\arrow[scale=1,>=stealth]{>}}},font=\scriptsize]{
		\draw[postaction=decorate] (0,1) -- (1,1);
		\draw[postaction=decorate] (1,1) -- (2,1);
		\draw[postaction=decorate] (1,0) -- (1,1);
		\draw[postaction=decorate] (1,1) -- (1,2);
		\node at (.4,1.6) {$z$};
	} \!
	& \color{gray!85}{0} & \color{gray!85}{0} & \color{gray!85}{0} \\ \color{gray!85}{0} &
	 \tikz[baseline={([yshift=-.5*10pt*0.6]current bounding box.center)},scale=0.3,decoration={markings, mark=at position 0.75 with {\arrow[scale=1,>=stealth]{>}}},font=\scriptsize]{
		\draw[postaction=decorate] (0,1) -- (1,1);
		\draw[postaction=decorate] (1,1) -- (2,1);			
		\draw[postaction=decorate] (1,1) -- (1,0);			
		\draw[postaction=decorate] (1,2) -- (1,1);
		\draw[white] (1,0) -- (1,-.2);
		\node at (.4,1.6) {$z$};
	}
	&
	\tikz[baseline={([yshift=-.5*10pt*0.6]current bounding box.center)},scale=0.3,decoration={markings, mark=at position 0.75 with {\arrow[scale=1,>=stealth]{>}}},font=\scriptsize]{
		\draw[postaction=decorate] (1,1) -- (0,1);
		\draw[postaction=decorate] (1,1) -- (2,1);
		\draw[postaction=decorate] (1,0) -- (1,1);
		\draw[postaction=decorate] (1,2) -- (1,1);
		\draw[white] (1,0) -- (1,-.3);
		\node at (.4,1.6) {$z$};
	} \!
	& \color{gray!85}{0} \\ \color{gray!85}{0} &
	\!\tikz[baseline={([yshift=-.5*10pt*0.6]current bounding box.center)},scale=0.3,decoration={markings, mark=at position 0.75 with {\arrow[scale=1,>=stealth]{>}}},font=\scriptsize]{
		\draw[postaction=decorate] (0,1) -- (1,1);
		\draw[postaction=decorate] (2,1) -- (1,1);
		\draw[postaction=decorate] (1,1) -- (1,0);
		\draw[postaction=decorate] (1,1) -- (1,2);
		\node at (.4,1.6) {$z$};
	}
	&
	\tikz[baseline={([yshift=-.5*10pt*0.6]current bounding box.center)},scale=0.3,decoration={markings, mark=at position 0.75 with {\arrow[scale=1,>=stealth]{>}}},font=\scriptsize]{
		\draw[postaction=decorate] (1,1) -- (0,1);
		\draw[postaction=decorate] (2,1) -- (1,1);
		\draw[postaction=decorate] (1,0) -- (1,1);
		\draw[postaction=decorate] (1,1) -- (1,2);
		\node at (.4,1.6) {$z$};
	} \!
	& \color{gray!85}{0} \\ \color{gray!85}{0} & \color{gray!85}{0} & \color{gray!85}{0} &
	 \tikz[baseline={([yshift=-.5*10pt*0.6]current bounding box.center)},scale=0.3,decoration={markings, mark=at position 0.75 with {\arrow[scale=1,>=stealth]{>}}},font=\scriptsize]{
		\draw[postaction=decorate] (1,1) -- (0,1);
		\draw[postaction=decorate] (2,1) -- (1,1);
		\draw[postaction=decorate] (1,1) -- (1,0);
		\draw[postaction=decorate] (1,2) -- (1,1);
		\node at (.4,1.6) {$z$};
	}
	\end{pmatrix} ,\label{eq:R-mat_dynamical}
	\end{gather}
where again $w=x_i -y_j$. In the diagrammatic version we have now explicitly oriented each line as indicated by a little arrow (not to be mistaken for a microscopic degree of freedom) at the `outgoing' end: this will help us keeping track of the flow of `time' (operator ordering) as we will have to rotate some of the figures in the presence of a reflecting end. For example,
	\begin{gather*}
	\tikz[baseline={([yshift=-.5*10pt*0.6]current bounding box.center)}, scale=0.6, font=\small, decoration={markings, mark=at position 0.625 with {\arrow[scale=1.25,>=stealth]{>}}}]{ % 0.59375 = 0.5+.75*1.25
		\draw[<-,postaction=decorate] (0,1) -- (1,1);
		\draw[postaction=decorate] (1,1) -- (2,1);
		\draw[postaction=decorate] (1,0) -- (1,1);
		\draw[->,postaction=decorate] (1,1) -- (1,2);
		\node at (.5,.5) {$z$};
	}
	 \overset{\curvearrowright}{=}
	\tikz[baseline={([yshift=-.5*10pt*0.6+.5pt]current bounding box.center)}, scale=0.6, font=\small, decoration={markings, mark=at position 0.625 with {\arrow[scale=1.25,>=stealth]{>}}}]{ % 0.59375 = 0.5+.75*1.25
		\draw[postaction=decorate] (0,1) -- (1,1);
		\draw[->,postaction=decorate] (1,1) -- (2,1);
		\draw[<-,postaction=decorate] (1,2) -- (1,1);
		\draw[postaction=decorate] (1,1) -- (1,0);
		\node at (.5,1.5) {$z$};
	}
 = b_+(w,z) .
	\end{gather*}
The interesting feature of this so-called `dynamical' $R$-matrix is that, unlike \eqref{eq:YBE}, the \emph{dynamical Yang--Baxter equation}
	\begin{gather} \label{eq:YBE_dynamical}
	\tikz[baseline={([yshift=-.5*10pt*0.6 +8.5pt]current bounding box.center)}, xscale=0.6, yscale=0.5, font=\footnotesize]{
		\pgfmathsetmacro{\csc}{1/sin(130)}
		\draw[->] (-130:1.5*\csc) node[below]{$i$} -- (50:1.5*\csc);
		\draw[->] (.5,-1.5) node[below]{$j$} -- (.5,1.5);
		\draw[->] (-50:1.5*\csc) node[below]{$k$} -- (130:1.5*\csc);
		\node at (180:.8) {$z$};
	}
	=
	\tikz[baseline={([yshift=-.5*10pt*0.6 +8.5pt]current bounding box.center)}, xscale=0.6, yscale=0.5, font=\footnotesize]{
		\pgfmathsetmacro{\csc}{1/sin(130)}
		\draw[->] (-130:1.5*\csc) node[below]{$i$} -- (50:1.5*\csc);
		\draw[->] (-.5,-1.5) node[below]{$j$} -- (-.5,1.5);
		\draw[->] (-50:1.5*\csc) node[below]{$k$} -- (130:1.5*\csc);
		\node at (180:1) {$z$};
	}
	\end{gather}
admits a block-diagonal solution~\eqref{eq:R-mat_dynamical} with an \emph{elliptic} parametrization for the weights:
	\begin{gather} \label{eq:vertices_dynamical}
	a_\pm(w,z) = [w+1] , \qquad b_\pm(w,z) = [w]\,\frac{[z\mp1]}{[z]} , \qquad c_\pm(w,z) = [1] \frac{[z\pm w]}{[z]} ,
	\end{gather}
where $[w] \coloneqq \E^{-\I \pi \tau/4} \vartheta_1(\gamma w;\tau)/2$ now denotes the odd Jacobi theta function with elliptic nome $\E^{\I \pi \tau} \in \mathbb{C}$ for $\Im(\tau)>0$, normalized such that we get the trigonometric case $[w] \to \sin(\gamma\,w)$ in the limit $\tau\to\I \infty$; if moreover $z\to\infty$ we recover \eqref{eq:vertices} up to some factors; by simply dropping all factors involving~$z$ we recover the setting of~\cite{GL_14}. The presence of an elliptic solution reflects the close connection between the elliptic \textsc{sos} model and the eight-vertex model.

One can proceed to define single-row operators similar to \eqref{eq:ABCD}, now depending on the height at any single face, that will obey relations akin to~\eqref{eq:RTT}. However, we also want to include the following.

{\bf Reflecting end.} A reflecting end is a special choice of boundary conditions that may be imposed at any of the four boundaries and is compatible with quantum integrability. The boundary is governed by the `dynamical $K$-matrix', of which we consider the diagonal case:
	\begin{gather} \label{eq:K-matrix_dynamical}
	K(x,z) = \begin{pmatrix} k_{+}(x,z) & \color{gray!85}{0} \\ \color{gray!85}{0} & k_{-}(x,z) \end{pmatrix} , \qquad
	\tikz[baseline={([yshift=-.5*10pt*0.6+3pt]current bounding box.center)}, font=\footnotesize, scale=0.6]{
		\draw[->] (-30:1.2) -- (0,0) -- (30:1.2);
		\fill[preaction={fill,white},pattern=north east lines, pattern color=gray] (0,-.75) rectangle (-.15,.75) ; \draw (0,-.75) -- (0,.75);
		\node at (-60:.8) {$z$};
	} =
	\begin{pmatrix} \tikz[baseline={([yshift=-.5*10pt*0.6+1pt]current bounding box.center)}, font=\scriptsize, scale=0.5,decoration={markings, mark=at position 0.65 with {\arrow[scale=1,>=stealth]{>}}}]{
		\draw[postaction=decorate] (-30:1.2) -- (0,0);
		\draw[postaction=decorate,->] (0,0) -- (30:1.2);
		\fill[preaction={fill,white},pattern=north east lines, pattern color=gray] (0,-.75) rectangle (-.15,.75) ; \draw (0,-.75) -- (0,.75);
		\node at (-60:1) {$z$};
		\node at (0:1.5) {$z-1$};
		\node at (60:1) {$z$};
	} & \color{gray!85}{0} \\ \color{gray!85}{0} & \tikz[baseline={([yshift=-.5*10pt*0.6+1pt]current bounding box.center)}, font=\scriptsize, scale=0.5,decoration={markings, mark=at position 0.65 with {\arrow[scale=1,>=stealth]{>}}}]{
		\draw[postaction=decorate] (0,0) -- (-30:1.2);
		\draw[postaction=decorate,<-] (30:1.2) -- (0,0);
		\fill[preaction={fill,white},pattern=north east lines, pattern color=gray] (0,-.75) rectangle (-.15,.75) ; \draw (0,-.75) -- (0,.75);
		\node at (-60:1) {$z$};
		\node at (0:1.5) {$z+1$};
		\node at (60:1) {$z$};
	} \end{pmatrix}	 .
	\end{gather}
At this point the orientation of the lines starts to matter: we will think of the diagram $\tikz[baseline={([yshift=-.5*10pt*0.6]current bounding box.center)}, font=\scriptsize, scale=0.3]{
	\draw (0,0) -- (-30:1.2);
	\draw[->] (0,0) -- (30:1.2);
	\fill[preaction={fill,white},pattern=north east lines, pattern color=gray] (0,-.75) rectangle (-.15,.75) ; \draw (0,-.75) -- (0,.75);
	% \node at (60:1) {$z$};
}$ as a (nearly) horizontal line, with associated parameter ${-x}$, that comes in at the bottom right and is reflected by the wall to continue to the top right with parameter ${+x}$.

In the diagrammatic notation, quantum integrability amounts to the possibility of sliding any line through crossings of any other two lines, see \eqref{eq:YBE}, \eqref{eq:RTT} and~\eqref{eq:YBE_dynamical}. From Cherednik~\cite{Che_84} we know that in the same spirit we should demand~\eqref{eq:K-matrix_dynamical} to obey the \emph{dynamical reflection equation}
	\begin{gather} \label{eq:reflection_dynamical}
	\begin{tikzpicture}[baseline={([yshift=-.5*10pt*0.75+5pt]current bounding box.center)}, font=\footnotesize, scale=0.75]
		\begin{scope}[shift={(0,-.5)}]
			\draw[->] (-50:1.25) -- (0,0) -- (50:2.75);
			\node at (-50:1.6) {$i'$};
			\node at (-70:1) {$z$};
		\end{scope}
		\begin{scope}[shift={(0,.5)}]
			\draw[->] (-15:2) -- (0,0) -- (15:2);
			\node at (-15:2.25) {$i$};
		\end{scope}
		\fill[preaction={fill,white},pattern=north east lines, pattern color=gray] (0,-1.75) rectangle (-.15,1.75); \draw (0,-1.75) -- (0,1.75);
	\end{tikzpicture}
	 =
	\begin{tikzpicture}[baseline={([yshift=-.5*10pt*0.75+5pt]current bounding box.center)}, font=\footnotesize, scale=0.75]
		\begin{scope}[shift={(0,.5)}]
			\draw[->] (-50:2.75) -- (0,0) -- (50:1.25);
			\node at (-50:3.1) {$i'$};
		\end{scope}
		\begin{scope}[shift={(0,-.5)}]
			\draw[->] (-15:2) -- (0,0) -- (15:2);
			\node at (-15:2.25) {$i$};
			\node at (-52.5:1) {$z$};
		\end{scope}
		\fill[preaction={fill,white},pattern=north east lines, pattern color=gray] (0,-1.75) rectangle (-.15,1.75); \draw (0,-1.75) -- (0,1.75);
	\end{tikzpicture} .
	\end{gather}
Note that this equation features (rotated versions of) the dynamical $R$-matrix. Given~\eqref{eq:R-mat_dynamical} the solution of the form~\eqref{eq:K-matrix_dynamical} involves a parameter $\kappa\in\mathbb{C}$ that we can associate to the wall:
	\begin{gather} \label{eq:vertices_dynamical_boundary}
	k_+(x,z) = [\kappa+x] \frac{[z+\kappa-x]}{[z+\kappa+x]} , \qquad k_-(x,z) = [\kappa-x] .
	\end{gather}
Because we consider `diagonal reflection' the value of $z$ is constant along the wall, cf.~\eqref{eq:K-matrix_dynamical}. It follows that, unlike in \eqref{eq:YBE_dynamical}, all $R$- and $K$-matrices in~\eqref{eq:reflection_dynamical} depend on the same value of the height in this case: the `diagonal' dynamical reflection equation reads
	\begin{gather*}
	R_{i,i'}(x_i-x_{i'},z) K_i(x_i,z) R_{i',i}(x_i+x_{i'},z) K_{i'}(x_{i'},z) \\
\qquad{}\times K_{i'}(x_{i'},z) R_{i,i'}(x_i+x_{i'},z) K_i(x_i,z) R_{i',i}(x_i-x_{i'},z) .
	\end{gather*}

Sklyanin~\cite{Skl_88} pioneered the algebraic formulation in the presence of a reflecting end. One should work with \emph{double}-row operators
	\begin{gather*} %\label{eq:ABCD_double}
	\begin{aligned}
	\mathcal{A}(x_i,z) & \coloneqq
	\tikz[baseline={([yshift=-.5*10pt*0.5+7.5pt]current bounding box.center)}, scale=0.5,font=\footnotesize,decoration={markings, mark=at position 0.65 with {\arrow[scale=1.25,>=stealth]{>}}}]{
		\draw[rounded corners=5pt] (0,0) -- (.75,-.5) -- (3.1,-.5) (3.9,-.5) -- (4.5,-.5); \draw[preaction=decorate] (5.5,-.5) -- (4.5,-.5);
		\draw[rounded corners=5pt] (0,0) -- (.75,+.5) -- (3.1,+.5) (3.9,+.5) -- (4.5,+.5); \draw[preaction=decorate,->] (4.5,+.5) -- (5.5,+.5);
		\foreach \y in {-1,...,1} {
			\draw (3.5+.2*\y,-.5) node{$\cdot\mathstrut$};
			\draw (3.5+.2*\y,+.5) node{$\cdot\mathstrut$};
		}
		\node at (.75,-1.1) {$z$};
		\fill[preaction={fill,white},pattern=north east lines, pattern color=gray] (0,-1.5) rectangle (-.15,1.5) ; \draw (0,-1.5) -- (0,1.5);
		\draw[->] (1.5,-1.5) node[below]{$1$} -- (1.5,1.5);
		\draw[->] (2.5,-1.5) node[below]{$2$} -- (2.5,1.5);
		\draw[->] (4.5,-1.5) node[below]{$L$} -- (4.5,1.5);
		\foreach \y in {-1,...,1} \draw (3.5+.2*\y,-1.9) node{$\cdot\mathstrut$};
	}
 , \qquad
	& \mathcal{B}(x_i,z) \coloneqq
	\tikz[baseline={([yshift=-.5*10pt*0.5+7.5pt]current bounding box.center)}, scale=0.5,font=\footnotesize,decoration={markings, mark=at position 0.65 with {\arrow[scale=1.25,>=stealth]{>}}}]{
		\draw[rounded corners=5pt] (0,0) -- (.75,-.5) -- (3.1,-.5) (3.9,-.5) -- (4.5,-.5); \draw[preaction=decorate] (4.5,-.5) -- (5.5,-.5);
		\draw[rounded corners=5pt] (0,0) -- (.75,+.5) -- (3.1,+.5) (3.9,+.5) -- (4.5,+.5); \draw[preaction=decorate,->] (4.5,+.5) -- (5.5,+.5);
		\foreach \y in {-1,...,1} {
			\draw (3.5+.2*\y,-.5) node{$\cdot\mathstrut$};
			\draw (3.5+.2*\y,+.5) node{$\cdot\mathstrut$};
		}
		\node at (.75,-1.1) {$z$};
		\fill[preaction={fill,white},pattern=north east lines, pattern color=gray] (0,-1.5) rectangle (-.15,1.5) ; \draw (0,-1.5) -- (0,1.5);
		\draw[->] (1.5,-1.5) node[below]{$1$} -- (1.5,1.5);
		\draw[->] (2.5,-1.5) node[below]{$2$} -- (2.5,1.5);
		\draw[->] (4.5,-1.5) node[below]{$L$} -- (4.5,1.5);
		\foreach \y in {-1,...,1} \draw (3.5+.2*\y,-1.9) node{$\cdot\mathstrut$};
	}
 , \\
	\mathcal{C}(x_i,z) & \coloneqq
	\tikz[baseline={([yshift=-.5*10pt*0.5+7.5pt]current bounding box.center)}, scale=0.5,font=\footnotesize,decoration={markings, mark=at position 0.65 with {\arrow[scale=1.25,>=stealth]{>}}}]{
		\draw[rounded corners=5pt] (0,0) -- (.75,-.5) -- (3.1,-.5) (3.9,-.5) -- (4.5,-.5); \draw[preaction=decorate] (5.5,-.5) -- (4.5,-.5);
		\draw[rounded corners=5pt] (0,0) -- (.75,+.5) -- (3.1,+.5) (3.9,+.5) -- (4.5,+.5); \draw[preaction=decorate,<-] (5.5,+.5) -- (4.5,+.5);
		\foreach \y in {-1,...,1} {
			\draw (3.5+.2*\y,-.5) node{$\cdot\mathstrut$};
			\draw (3.5+.2*\y,+.5) node{$\cdot\mathstrut$};
		}
		\node at (.75,-1.1) {$z$};
		\fill[preaction={fill,white},pattern=north east lines, pattern color=gray] (0,-1.5) rectangle (-.15,1.5) ; \draw (0,-1.5) -- (0,1.5);
		\draw[->] (1.5,-1.5) node[below]{$1$} -- (1.5,1.5);
		\draw[->] (2.5,-1.5) node[below]{$2$} -- (2.5,1.5);
		\draw[->] (4.5,-1.5) node[below]{$L$} -- (4.5,1.5);
		\foreach \y in {-1,...,1} \draw (3.5+.2*\y,-1.9) node{$\cdot\mathstrut$};
	}
 , \qquad
	& \mathcal{D}(x_i,z) \coloneqq
	\tikz[baseline={([yshift=-.5*10pt*0.5+7.5pt]current bounding box.center)}, scale=0.5,font=\footnotesize,decoration={markings, mark=at position 0.65 with {\arrow[scale=1.25,>=stealth]{>}}}]{
		\draw[rounded corners=5pt] (0,0) -- (.75,-.5) -- (3.1,-.5) (3.9,-.5) -- (4.5,-.5); \draw[preaction=decorate] (4.5,-.5) -- (5.5,-.5);
		\draw[rounded corners=5pt] (0,0) -- (.75,+.5) -- (3.1,+.5) (3.9,+.5) -- (4.5,+.5); \draw[preaction=decorate,<-] (5.5,+.5) -- (4.5,+.5);
		\foreach \y in {-1,...,1} {
			\draw (3.5+.2*\y,-.5) node{$\cdot\mathstrut$};
			\draw (3.5+.2*\y,+.5) node{$\cdot\mathstrut$};
		}
		\node at (.75,-1.1) {$z$};
		\fill[preaction={fill,white},pattern=north east lines, pattern color=gray] (0,-1.5) rectangle (-.15,1.5) ; \draw (0,-1.5) -- (0,1.5);
		\draw[->] (1.5,-1.5) node[below]{$1$} -- (1.5,1.5);
		\draw[->] (2.5,-1.5) node[below]{$2$} -- (2.5,1.5);
		\draw[->] (4.5,-1.5) node[below]{$L$} -- (4.5,1.5);
		\foreach \y in {-1,...,1} \draw (3.5+.2*\y,-1.9) node{$\cdot\mathstrut$};
	}
 .
	\end{aligned}
	\end{gather*}
By virtue of \eqref{eq:YBE_dynamical} and~\eqref{eq:reflection_dynamical} these operators also obey certain relations, contained in the dyna\-mi\-cal double-row analogue of~\eqref{eq:RTT}:
	\begin{gather} \label{eq:ref_alg}
	\tikz[baseline={([yshift=-.5*10pt*0.5+7.5pt]current bounding box.center)}, scale=0.5,font=\footnotesize]{
		\begin{scope}[shift={(0,1)}]
			\draw[rounded corners=5pt] (0,0) -- (.75,-.5) -- (3.1,-.5) (3.9,-.5) -- (5.5,-.5) -- (6.5,-1.5) -- (8,-1.5) node[right]{$i$};
			\draw[rounded corners=5pt,->] (0,0) -- (.75,+.5) -- (3.1,+.5) (3.9,+.5) -- (8,+.5);
			\foreach \y in {-1,...,1} {
				\draw (3.5+.2*\y,-.5) node{$\cdot\mathstrut$};
				\draw (3.5+.2*\y,+.5) node{$\cdot\mathstrut$};
			}
		\end{scope}
		\begin{scope}[shift={(0,-1)}]
			\draw[rounded corners=5pt] (0,0) -- (.75,-.5) -- (3.1,-.5) (3.9,-.5) -- (8,-.5) node[right]{$i'$};
			\draw[rounded corners=5pt,->] (0,0) -- (.75,+.5) -- (3.1,+.5) (3.9,+.5) -- (5.5,+.5) -- (7.5,3.5) -- (8,3.5);
			\foreach \y in {-1,...,1} {
				\draw (3.5+.2*\y,-.5) node{$\cdot\mathstrut$};
				\draw (3.5+.2*\y,+.5) node{$\cdot\mathstrut$};
			}
		\end{scope}
		\node at (.75,-2.1) {$z$};
		\fill[preaction={fill,white},pattern=north east lines, pattern color=gray] (0,-2.5) rectangle (-.15,3.5) ; \draw (0,-2.5) -- (0,3.5);
		\draw[->] (1.5,-2.5) node[below]{$1$} -- (1.5,3.5);
		\draw[->] (2.5,-2.5) node[below]{$2$} -- (2.5,3.5);
		\draw[->] (4.5,-2.5) node[below]{$L$} -- (4.5,3.5);
		\foreach \y in {-1,...,1} \draw (3.5+.2*\y,-2.9) node{$\cdot\mathstrut$};
	} =
	\tikz[baseline={([yshift=-.5*10pt*0.5+7.5pt]current bounding box.center)}, scale=0.5,font=\footnotesize]{
		\begin{scope}[shift={(0,-1)}]
			\draw[rounded corners=5pt] (0,0) -- (.75,-.5) -- (3.1,-.5) (3.9,-.5) -- (8,-.5) node[right]{$i$};
			\draw[rounded corners=5pt,->] (0,0) -- (.75,+.5) -- (3.1,+.5) (3.9,+.5) -- (5.5,+.5) -- (6.5,+1.5) -- (8,+1.5);
			\foreach \y in {-1,...,1} {
				\draw (3.5+.2*\y,-.5) node{$\cdot\mathstrut$};
				\draw (3.5+.2*\y,+.5) node{$\cdot\mathstrut$};
			}
		\end{scope}
		\begin{scope}[shift={(0,1)}]
			\draw[rounded corners=5pt] (0,0) -- (.75,-.5) -- (3.1,-.5) (3.9,-.5) -- (5.5,-.5) -- (7.5,-3.5) -- (8,-3.5) node[right]{$i'$};
			\draw[rounded corners=5pt,->] (0,0) -- (.75,+.5) -- (3.1,+.5) (3.9,+.5) -- (8,+.5);
			\foreach \y in {-1,...,1} {
				\draw (3.5+.2*\y,-.5) node{$\cdot\mathstrut$};
				\draw (3.5+.2*\y,+.5) node{$\cdot\mathstrut$};
			}
		\end{scope}
		\node at (.75,-2.6) {$z$};
		\fill[preaction={fill,white},pattern=north east lines, pattern color=gray] (0,-3.5) rectangle (-.15,2.5) ; \draw (0,-3.5) -- (0,2.5);
		\draw[->] (1.5,-3.5) node[below]{$1$} -- (1.5,2.5);
		\draw[->] (2.5,-3.5) node[below]{$2$} -- (2.5,2.5);
		\draw[->] (4.5,-3.5) node[below]{$L$} -- (4.5,2.5);
		\foreach \y in {-1,...,1} \draw (3.5+.2*\y,-3.9) node{$\cdot\mathstrut$};
	} .
	\end{gather}
The explicit commutation rules are again found by fixing the microscopic degrees of freedom on the four external horizontal edges, and the structure constants of the resulting \emph{dynamical reflection algebra} are built from the entries of the dynamical $R$-matrix. Just as for the $K$-mat\-rices in~\eqref{eq:reflection_dynamical}, the double-row operators in~\eqref{eq:ref_alg} depend on the same value of the height for diagonal reflection.

With all these preliminaries in place the reflecting-end partition function, which equals \eqref{eq:partition_function_general_dynamical} times the boundary weights, can be defined as
	\begin{gather} \label{eq:partition_function_SOS}
	\mathcal{Z}(\vec{x}) \coloneqq \bra{{\color{Red}\Down\cdots\Down}} \mathcal{B}(x_1,z) \cdots \mathcal{B}(x_L,z) \ket{{\color{Cerulean}\Up\cdots\Up}} = \tikz[baseline={([yshift=-.5*10pt*0.5]current bounding box.center)}, font=\footnotesize, scale=0.5, decoration={markings, mark=at position 0.65 with {\arrow[scale=1.25,>=stealth]{>}}}]{% 0.59375 = 0.5+.75*1.25
		\foreach \y in {1,2,3.5} {
			\draw[rounded corners=5pt] (0,1.5*\y-.25) -- +(.75,-.35) -- +(3.1,-.35) +(3.9,-.35) -- +(4.5,-.35);
			\foreach \yy in {-1,...,1} \draw (3.5+.2*\yy,1.5*\y-.25-.35) node{$\cdot\mathstrut$};
			\draw[postaction=decorate] (4.5,1.5*\y-.25-.35) -- +(1,0);
			\draw[rounded corners=5pt] (0,1.5*\y-.25) -- +(.75,+.35) -- +(3.1,+.35) +(3.9,+.35) -- +(4.5,+.35);
			\foreach \yy in {-1,...,1} \draw (3.5+.2*\yy,1.5*\y-.25+.35) node{$\cdot\mathstrut$};
			\draw[postaction=decorate,->] (4.5,1.5*\y-.25+.35) -- +(1,0);
		}		
		\fill[preaction={fill,white},pattern=north east lines, pattern color=gray] (0,0) rectangle (-.15,1.5*3.5-.25+.35+.9) ; \draw (0,0) -- (0,1.5*3.5-.25+.35+.9);
		\foreach \x in {1,2,4} {
			\draw[postaction=decorate,color=Cerulean] (\x+.5,0) -- +(0,.9);
			\draw (\x+.5,.9) -- (\x+.5,3.475) (\x+.5,4.275) -- (\x+.5,1.5*3.5-.25+.35);
			\foreach \y in {-1,...,1} \draw (\x+.5,3.875+.2*\y) node{$\cdot\mathstrut$};
			\draw[<-] (\x+.5,1.5*3.5-.25+.35+.9) -- +(0,-.1);
			\draw[postaction=decorate,color=Red] (\x+.5,1.5*3.5-.25+.35+.9) -- +(0,-.9);
		}
		\node at (.75,.4) {$z$};
	} .
	\end{gather}
There are $2L$ horizontal lines, pairwise connected by the reflecting end, and $L$ vertical lines. The domain walls on the three other ends look just as in \eqref{eq:partition_function}. The result is consistent with the ice rule: there are equally many arrows pointing in and out of the lattice. Again, due to the diagonal reflection~\eqref{eq:K-matrix_dynamical} each face along the wall has the same height~$z$, and all~$\mathcal{B}$s have equal `dynamical' argument. This is why we choose to suppress the dependence on~$z$ in our notation. Since~\eqref{eq:partition_function_SOS} is a polynomial in the weights~\eqref{eq:vertices_dynamical} and~\eqref{eq:vertices_dynamical_boundary}, the reflecting-end partition function is an elliptic polynomial, or more precisely a `higher-order theta function', in all parameters.

{\bf Korepin--Izergin method.} The reflecting-end partition function was computed using the approach from Section~\ref{s:KI} for the (nondynamical, trigonometric) six-vertex model by Tsu\-chiya~\cite{Tsu_98} and in the (trigonometric and elliptic) \textsc{sos} setting by~Filali and Kitanine~\cite{Fil_11b,Fil_11a, FK_10}.

The analytic properties are best formulated in terms of the `renormalized' partition function
	\begin{gather} \label{eq:Zbar}
	\bar{\mathcal{Z}}(\vec{x};\vec{y}\,) \coloneqq \prod_{i=1}^L \frac{[z+\kappa+x_i]}{[2x_i]} \times \mathcal{Z}(\vec{x};\vec{y}\,) .
	\end{gather}
Namely,
\begin{itemize}\itemsep=0pt
	\item $\mathcal{Z}$ (whence $\bar{\mathcal{Z}}$) is doubly symmetric;
	\item $\mathcal{Z}$ is crossing symmetric: for any $i$, $\bar{\mathcal{Z}}$ is invariant under $x_i\mapsto {-x_i} - 1$;
	\item $\bar{\mathcal{Z}}_L$ is an elliptic polynomial of degree $2(L-1)$ in each variable $x_i$, i.e., a higher-order theta function of order $2(L-1)$ and norm $(L-1)\gamma$.
\end{itemize}
Thus, the polynomial degree is essentially twice that of the domain-wall partition function. This doubling can be understood as a consequence of the crossing symmetry, which in turn is due to the double-row structure and the reflecting end. As a consequence, longer and more complicated expressions than those in Sections~\ref{s:setup} and~\ref{s:constr} are unavoidable. Let us again use the shorthand~\eqref{eq:shorthand}.

The recurrence relations are again of the form \eqref{eq:Korepin_derivation}, \eqref{eq:Korepin_alt}. Explicitly,
	\begin{gather}
 \mathcal{Z}_L(\vec{x};\vec{y}\,)\big|_{x_L = \pm y_L} = \mathcal{Z}_{L-1}(x_1,\dots,x_{L-1};y_1,\dots,y_{L-1}) \nonumber\\
\hphantom{\mathcal{Z}_L(\vec{x};\vec{y}\,)\big|_{x_L = \pm y_L} =}{} \times k_\mp(\pm y_L,z) [1 ,\pm 2y_L] \frac{[z\pm(L-1)-1]}{[z\pm(L-1)]}\nonumber \\
\hphantom{\mathcal{Z}_L(\vec{x};\vec{y}\,)\big|_{x_L = \pm y_L} =}{} \times \prod_{i=1}^{L-1} [x_i \mp y_L+1 ,x_i\pm y_L ,\pm y_L \pm y_i+1 ,\pm y_L - y_i+1] \nonumber \\
\hphantom{\mathcal{Z}_L(\vec{x};\vec{y}\,)\big|_{x_L = \pm y_L} =}{}\times \frac{[z\pm(2i-L-1)-1]}{[z\pm(2i-L-1)]} .\label{eq:Tsu_Fil_Kit}
	\end{gather}
For brevity we have given the crossing-symmetric version of \eqref{eq:Korepin_alt}. By the double symmetry and crossing symmetry these give enough equations to uniquely characterize the reflecting-end partition function. To solve this using Lagrange interpolation one employs the basis of~$2L$ elliptic polynomials of degree~$2L-1$ such that for each point $\pm y_j$ only one is nonzero.

Tsuchiya, Filali and Kitanine managed to find the answer in terms of a determinant:
	\begin{gather}
	\mathcal{Z}_L(\vec{x};\vec{y}\,) = \prod_{i=1}^L [\kappa-y_i , 2x_i] \frac{[z+\kappa+y_i , z+(2i-L-2)]}{[z+\kappa+x_i , z+(L-i)]} \nonumber\\
 \hphantom{\mathcal{Z}_L(\vec{x};\vec{y}\,) =}{} \times [1]^L \frac{\prod\limits_{i,j=1}^L [x_i-y_j+1 , x_i-y_j , x_i+y_j+1 , x_i+y_j]}{\prod\limits_{i<j}^L [x_i+x_j+1 ,x_i-x_j ,y_j+y_i ,y_j-y_i]} \nonumber\\
\hphantom{\mathcal{Z}_L(\vec{x};\vec{y}\,) =}{} \times \det_{i,j}\left(\frac{1}{[x_i-y_j+1 , x_i-y_j , x_i+y_j+1 , x_i+y_j]}\right) . \label{eq:Tsu_Fil_Kit_det}
	\end{gather}
Surprisingly, both the boundary parameter~$\kappa$ and the dynamical parameter~$z$ appear exclusively in the prefactor on the first line. Also note that the factors in this first line that depend on $x_i$ are precisely those removed in~\eqref{eq:Zbar}. The third line of~\eqref{eq:Tsu_Fil_Kit_det} is just the crossing-symmetric extension of Izergin's determinant~\eqref{eq:Izergin}.

{\bf Functional method.} The present setting contains several layers of complexity compared to Section~\ref{s:setup}. Let us sketch how the functional method can be adapted. Expression \eqref{eq:partition_function_SOS} gives us ingredient~\eqref{it:alg_expr} from Section~\ref{s:applicability}. For \eqref{it:suitable_op} a candidate is $\mathcal{A}$; by the ice rule it obeys~\mbox{(\ref{it:suitable_op}\textsuperscript{\,\ref{it:op_eigenvector}})}. The reflecting end, however, makes the relevant commutation rule more complicated than \eqref{eq:AB}:
\begin{subequations} \label{eq:AB_DB_double}
	\begin{gather}
	\mathcal{A}(x,z) \mathcal{B}(x',z) = \substack{\text{some}\\\text{structure}\\\text{constant}} \times \mathcal{B}(x',z) \mathcal{A}(x,z) + \substack{\text{another} \\ \text{structure}\\ \text{constant}} \times \mathcal{B}(x,z) \mathcal{A}(x',z)\nonumber\\
\hphantom{\mathcal{A}(x,z) \mathcal{B}(x',z) = \substack{\text{some}\\\text{structure}\\\text{constant}} \times \mathcal{B}(x',z) \mathcal{A}(x,z)}{}
 + \substack{\text{and another} \\ \text{structure}\\\text{constant}} \times \mathcal{B}(x,z) \mathcal{D}(x',z)\label{eq:AB_double}
	\end{gather}
contains a term where the $\mathcal{A}$ has not only swapped parameters with $\mathcal{B}$ but moreover turned into a $\mathcal{D}$. Requirement~\mbox{(\ref{it:suitable_op}\textsuperscript{\,\ref{it:suitable_op_b}})} is safeguarded by another commutation rule contained in \eqref{eq:ref_alg} that allows one to move the $\mathcal{D}$ to the right through the remaining $\mathcal{B}$s:
	\begin{gather}
	\mathcal{D}(x,z) \mathcal{B}(x',z) = \substack{\text{some}\\\text{structure}\\\text{constant}} \times \mathcal{B}(x',z) \mathcal{D}(x,z) + \substack{\text{another} \\ \text{structure}\\\text{constant}} \times \mathcal{B}(x,z) \mathcal{D}(x',z) \nonumber\\
\hphantom{\mathcal{D}(x,z) \mathcal{B}(x',z) =}{} + \substack{\text{yet another} \\ \text{structure}\\\text{constant}} \times \mathcal{B}(x',z) \mathcal{A}(x,z) + \substack{\text{and another} \\ \text{structure}\\\text{constant}} \times \mathcal{B}(x,z) \mathcal{A}(x',z) .\label{eq:DB_double}
	\end{gather}
\end{subequations}
(The structure constants differ from \eqref{eq:AB_double}.) By the ice rule $\ket{{\color{Cerulean}\Up\cdots\Up}}$ is an eigenvector for $\mathcal{D}$ too, so condition \eqref{it:suitable_op} is met and one can derive a functional equation for the partition function.

From \eqref{eq:AB_DB_double} we will obtain something of the form \eqref{eq:derivation_3}, where $\textcolor{Cerulean}{\text{eigenvalue}_i}$ will contain a contribution from $\mathcal{A}$ and one from $\mathcal{D}$. Fairly compact expressions, see~\eqref{**} below, can be found for these eigenvalues with some tricks~\cite{Skl_88}. For the computation of the coefficients in \eqref{eq:derivation_3} one can again exploit that the $\mathcal{B}$s commute amongst themselves, cf.~\eqref{it:comm_symm}, provided one brings~\eqref{eq:AB_DB_double} to a simpler, more symmetric form following Sklyanin~\cite{Skl_88}. Indeed, $\mathcal{D}$ can be replaced with a linear combination $\tilde{\mathcal{D}}=\mathcal{D}+\text{factor}\times\mathcal{A}$ where the factor is chosen such that the double-row operators $\mathcal{A}$, $\mathcal{B}$ and $\tilde{\mathcal{D}}$ obey relations like \eqref{eq:AB_DB_double} but without the last term in \eqref{eq:DB_double}.

The result is a linear functional equation for the partition function~\eqref{eq:partition_function_SOS} with the same form as~\eqref{eq:functional}. The coefficient for $\nu=0$ has the same structure as in \eqref{eq:M_0}:
\begin{subequations} \label{eq:M_0i_dyn_refl}
	\begin{gather}
	M_0(x_0;\vec{x}\,) \coloneqq \Lambda^{{\color{Red}\Down \cdots\Down}}_{\mathcal{A}(x_0)} - \Lambda^{{\color{Cerulean}\Up \cdots\Up}}_{\mathcal{A}(x_0)} \prod_{j=1}^{L} \frac{[x_j-x_0+1 , x_0+x_j]}{[x_j-x_0 , x_0+x_j+1]} , % \label{eq:M_0}
\end{gather}
where the eigenvalues $\Lambda_{\mathcal{A}}$ will be given shortly. Due to the $\tilde{\mathcal{D}}$ picked up along the way the coefficients for $1\leq i \leq L$ now consist of two terms, each with the structure of~\eqref{eq:M_i}:
	\begin{gather}
	 M_i(x_0;\vec{x}\,) \coloneqq \Lambda^{{\color{Cerulean}\Up \cdots\Up}}_{\mathcal{A}(x_i)} \frac{[1 , 2x_i , x_0-x_i-z+(L-1)]}{[x_0-x_i , 2x_i+1 , z+(L-1)]} \prod_{\substack{j=1 \\j \neq i}}^{L} \frac{[x_j-x_i+1 , x_i+x_j]}{[x_j-x_i , x_i+x_j+1]}
\label{eq:M_i_dyn_refl}\\
\hphantom{M_i(x_0;\vec{x}\,) \coloneqq}{} + \Lambda^{{\color{Cerulean}\Up\,\cdots\Up}}_{\tilde{\mathcal{D}}(x_i)} \frac{[1 , z+(L-2)-x_0-x_i , z-(L-1)]}{[x_0+x_i+1 , z+(L-1) , z-L]} \prod_{\substack{j=1 \\ j \neq i}}^{L} \frac{[x_i-x_j+1 , x_i+x_j+2]}{[x_i-x_j , x_i+x_j+1]} .
\nonumber
\end{gather}
The calculation of the eigenvalues, using yet another trick due to Sklyanin~\cite{Skl_88} to get relatively compact expressions, gives
	\begin{gather}
	\Lambda^{{\color{Cerulean}\Up \cdots\Up}}_{\mathcal{A}(x)} = [\kappa+x] \frac{[z+\kappa-x]}{[z+\kappa+x]} \prod_{j=1}^L [x-y_j+1 , x+y_j+1] , \nonumber\\
	\Lambda^{{\color{Cerulean}\Up \cdots\Up}}_{\widetilde{\mathcal{D}}(x)} = [\kappa-x-1] \frac{[2x , z+\kappa+x+1 , z-L]}{[2x+1 , z+\kappa+x , z-(L-1)]} \prod_{j=1}^L [x-y_j , x+y_j] , \nonumber\\
	\Lambda^{{\color{Red}\Down \cdots\Down}}_{\mathcal{A}(x)} = [\kappa-x] \frac{[1 , z+(L-1)-2x]}{[2x+1 , z+(L-1)]} \prod_{j=1}^L [x-y_j+1 , x+y_j+1] \nonumber\\
\hphantom{\Lambda^{{\color{Red}\Down \cdots\Down}}_{\mathcal{A}(x)} = }{}	 + [\kappa+x+1] \frac{[2x , z+\kappa-x-1 , z+L]}{[2x+1 , z+\kappa+x , z+(L-1)]} \prod_{j=1}^L [x-y_j , x+y_j] ,\label{**}
	\end{gather}
\end{subequations}

The analysis of the functional equation is similar to that in Section~\ref{s:analysis}, though due to the added complexity we did not manage to make each step rigorous. As before the equation immediately reveals several analytic properties of its solutions: any reasonable (viz.~meromorphic) solution is a symmetric function of the $x_i$, is crossing symmetric, and ought to have the same (elliptic) polynomial structure as the reflecting-end partition function.\footnote{The coefficients of the functional equation are ratios of elliptic polynomials. A comparison of their orders and norms give strong evidence for what polynomial structure the solution will have. (To turn the argument of \cite{Lam_16} into a proof one has to make sure that there cannot be any cancellations between the different terms.)} Because of the crossing symmetric there should be more `special zeroes' with values that are readily guessed; we did not prove ingredient~\eqref{it:special_zeroes} in \cite{Lam_15,Lam_16} but confirmed it numerically for $L\leq 15$. Using these special zeroes one can recover the recurrence relations~\eqref{eq:Tsu_Fil_Kit}~\cite{Lam_16}. For the reduction step~\eqref{it:reduction} a proof is lacking in \cite{Lam_15,Lam_16}, but numerical checks have been performed for $L\leq12$~\cite{Lam_15}. Despite these gaps in the rigorous treatment of the functional method in this case one can still put everything together to obtain a recipe leading to a formula for a solution, or \emph{the} solution if one believes the numerical checks. The result is again a symmetrized sum:
\begin{subequations} \label{eq:symmetrized_sum_dyn}
	\begin{gather}
	\mathcal{Z}(\vec{x};\vec{y}\,) = \prod_{i=1}^L [\kappa-y_i , 2x_i] \frac{[z+\kappa+y_i ,z+(2i-L-2)]}{[z+\kappa+x_i ,z+(L-i)]} \nonumber \\
	\hphantom{\mathcal{Z}(\vec{x};\vec{y}\,) =}{} \times [1]^L \sum_{\sigma \in S_L} \prod_{n=1}^{L} m_n(x_{\sigma 1},x_{\sigma 2},\dots,x_{\sigma n}) \prod_{i < j}^L \frac{[ x_{\sigma i} - y_j , x_{\sigma i} + y_j + 1 ]}{[x_{\sigma i}-x_{\sigma j} , x_{\sigma i}+x_{\sigma j}+1]} ,\label{eq:symmetrized_sum_dyn_a}
\end{gather}
where, to match the first line of \eqref{eq:Tsu_Fil_Kit_det}, compared to \cite{GL_14,Lam_15,Lam_16} we have moved some factors to
\begin{gather}
	m_n(x_1,\dots,x_n) \coloneqq \frac{[\kappa+x_n , x_n+y_n+1 ,z+\kappa-x_n , z+ n +x_n-y_n]}{[\kappa+y_n , 2x_n + 1 , z+\kappa-y_n , z + n ] } \nonumber \\
\hphantom{m_n(x_1,\dots,x_n) \coloneqq}{} \times \prod_{j=1}^{n-1} [x_n-y_j+1 , x_n+y_j+1 , x_j-x_n+1 , x_j+x_n] \nonumber \\
\hphantom{m_n(x_1,\dots,x_n) \coloneqq}{}
 + (-1)^{n-1} \frac{[\kappa-x_n-1 , x_n-y_n , z+\kappa+x_n+1 , z+(n-1)-x_n-y_n] }{[\kappa+y_n , 2x_n + 1 , z+\kappa-y_n , z+n] } \nonumber \\
\hphantom{m_n(x_1,\dots,x_n) \coloneqq}{} \times \prod_{j=1}^{n-1} [x_n-y_j , x_n+y_j , x_n-x_j+1 , x_n+x_j+2] .\label{eq:m_n}
	\end{gather}
\end{subequations}
To get some feeling for this result we examine analytic structure of~\eqref{eq:symmetrized_sum_dyn}. For the first line the zero at $x_i=0$ and the simple pole at $x_i = {-z}-\kappa$ were already anticipated in~\eqref{eq:Zbar}. The simple poles on the second line are removable: for $x_i = x_j$ the argument is as for~\eqref{eq:symmetrized_sum}, while crossing symmetry guarantees that the poles at $x_i= {-x_j}-1$ are also removable. The poles of the~$m_n$ at the fixed point under crossing symmetry, $x_i = -1/2$, are removable too.

\subsection{A new expression for the reflecting-end partition function} \label{s:new}

Let us now show how the results of \cite{GL_14,Lam_15,Lam_16} may be improved by rewriting \eqref{eq:symmetrized_sum_dyn} in a more appealing way. Focus on the quantities~\eqref{eq:m_n}, which form the major complication with respect to~\eqref{eq:symmetrized_sum}. The $n$th such factor in \eqref{eq:symmetrized_sum_dyn_a} essentially comes from the~$(L-n)$th iteration of the recursive recipe obtained from the functional equation, cf.~\eqref{eq:F_via_mn}:
	\begin{gather*}
	m_n(x_1,\dots,x_n) = \frac{1}{[\kappa+y_n , 2\,x_n+1,2x_n , 1]} \frac{[z+\kappa+x_n , z+(n-1) ]}{[z+\kappa-y_n, z+n]} M_n(y_n-1;x_1,\dots,x_n) .
	\end{gather*}
We can thus trace this complexity back the presence of the second term in~\eqref{eq:M_i_dyn_refl} and hence to the additional term in \eqref{eq:AB_double}, i.e., to the reflection algebra itself. This is reflected in the crossing invariance of~\eqref{eq:m_n}: both terms in \eqref{eq:m_n} are clearly separately invariant under $x_j \mapsto {-x_j} -1$ for $j<n$, while crossing $x_n \mapsto {-x_n} -1$ exchanges the two terms in \eqref{eq:m_n}. For $I\subseteq\{1,\dots,L\}$ define $r=r_I$ to act on the spectral parameters by $x_i^r \coloneqq {-x_i}-1$ if $i\in I$, $x_i^r \coloneqq x_i$ else, and write $R_L \coloneqq \{ r_I \mid I \subseteq \{1,\dots,L\} \, \} \cong (\mathbf{Z}_2)^L$ for the group of all such `reflections', cf.~\cite{Fra_17}. By the preceding observation we can rewrite
	\begin{gather*}
	\prod_{n=1}^L m_n(x_1,\dots,x_n) = \sum_{r\in R_L} \prod_{i=1}^L \frac{[\kappa+x_i^r , x_i^r+y_i +1 , z+\kappa-x_i^r , z + i + x_i^r - y_i] }{[\kappa+y_i , 2x_i^r + 1 , z+\kappa-y_i , z + i ] } \\
\hphantom{\prod_{n=1}^L m_n(x_1,\dots,x_n) =}{} \times \prod_{i<j}^L \big[x_j^r-y_i+1 , x_j^r+y_i+1 , x_i^r-x_j^r+1 , x_i^r+x_j^r\big] .
	\end{gather*}
With a little algebra our previous result~\eqref{eq:symmetrized_sum_dyn} can now be brought to a \emph{crossing}-symmetrized sum with $2^L$ terms:
\begin{subequations} \label{eq:new_formula_SOS}
	\begin{gather}
 \mathcal{Z}(\vec{x};\vec{y}\,) = \prod_{i=1}^L [\kappa-y_i , 2x_i] \frac{[z+\kappa+y_i ,z+(2i-L-2)]}{[z+\kappa+x_i ,z+(L-i)]} \nonumber\\
\hphantom{\mathcal{Z}(\vec{x};\vec{y}\,) =}{} \times \sum_{r\in R_L} \prod_{i=1}^L \frac{[\kappa+x_i^r , z+\kappa-x_i^r] }{[ \kappa+y_i , 2x_i^r +1 , z+\kappa-y_i] } \prod_{i<j}^L \frac{[x_i^r+x_j^r]}{[x_i^r+x_j^r+1]}
 \nonumber\\
\hphantom{\mathcal{Z}(\vec{x};\vec{y}\,) =}{} \times
 \prod_{i,j=1}^L [x_i^r + y_j + 1] \times Z_\text{ell}(\vec{x}\,{}^r;\vec{y}\,) .\label{eq:crossing_symmetrized_sum_dyn}
\end{gather}
We have kept the first line intact, and the second and third lines are manifestly invariant under crossing, cf.~the property of~\eqref{eq:Zbar}. The explicit factors in this second line are symmetric in the $x_i$, which enabled us to take them out the sum over $\sigma\in S_L$ from \eqref{eq:symmetrized_sum_dyn}. Finally,
\begin{gather}	\label{eq:partition_function_SOS_dwbc}
 Z_\text{ell}(\vec{x};\vec{y}\,) = [1]^L \sum_{\sigma\in S_L} \prod_{i=1}^L \frac{[z+ i+x_{\sigma i}-y_i]}{[z+i]} \prod_{i<j}^L [x_{\sigma i} -y_j, x_{\sigma j} -y_i +1 ] \frac{[x_{\sigma i} -x_{\sigma j} +1]}{[x_{\sigma i} - x_{\sigma j}]}\!\!\!\!\!
	\end{gather}
\end{subequations}
is the partition function of the elliptic \textsc{sos} model with domain-wall boundaries as in \eqref{eq:partition_function}. The symmetrized sum~\eqref{eq:partition_function_SOS_dwbc}, cf.~\eqref{eq:symmetrized_sum}, first appeared as an elliptic weight function~\cite{TV_97}, so \eqref{eq:partition_function_SOS_dwbc} is some sort of relation between elliptic weight functions of type $BC$, with Weyl group $S_L\ltimes R_L$, and type $A$. Note that the elliptic domain-wall partition function can be characterized like in Section~\ref{s:KI} \cite{PRS_08, Ros_09} and admits an expression as a \emph{sum} of determinants~\cite{Ros_09}.
	
By \eqref{eq:Tsu_Fil_Kit_det} the sum over $r\in R_L$ as written in~\eqref{eq:crossing_symmetrized_sum_dyn} is actually independent of $\kappa$ and~$z$, even though its summands are not. One should be able to see this remarkable cancellation, and explicitly recover the determinant~\eqref{eq:Tsu_Fil_Kit_det}, from~\eqref{eq:crossing_symmetrized_sum_dyn} by an elliptic version of Langrange interpolation.

Setting $\sgn r_I = (-1)^{\#I}$ for $r_I \in R_L$, so that $\prod\limits_{i=1}^L [2x_i^r+1] = \sgn(r) \prod\limits_{i=1}^L [2x_i+1]$, our formula~\eqref{eq:partition_function_SOS_dwbc} may be recast in the slightly more elegant form
	\begin{gather}
	 \mathcal{Z}(\vec{x};\vec{y}\,) = \prod_{i=1}^L \frac{[\kappa-y_i , 2x_i , z+\kappa+y_i \,,z+(2i-L-2)]}{[\kappa+y_i , 2x_i +1 , z+\kappa+x_i ,z+(L-i)]} \sum_{r\in R_L} \sgn(r) \prod_{i=1}^L [\kappa+x_i^r] \frac{[z+\kappa-x_i^r]}{[z+\kappa-y_i]} \nonumber\\
\hphantom{\mathcal{Z}(\vec{x};\vec{y}\,)=}{}	 \times \prod_{i<j}^L \frac{[x_i^r+x_j^r]}{[x_i^r+x_j^r+1]} \prod_{i,j=1}^L [x_i^r + y_j + 1] \times Z_\text{ell}(\vec{x}\,{}^r;\vec{y}\,) .\label{eq:new_formula_SOS_rewrite}
	\end{gather}
For the degenerate case of the six-vertex model, $[w]\to\sin(\gamma\, w)$, we remove all factors featuring~$z$ to obtain an expression for the ordinary reflecting-end partition function~\cite{GL_14, Tsu_98} in terms of the domain-wall partition function~\eqref{eq:Izergin}, \eqref{eq:symmetrized_sum}:
	\begin{gather}
	\mathcal{Z}_\text{tri}(\vec{x};\vec{y}\,)
	= \prod_{i=1}^L \frac{k_-(y_i) b(2x_i)}{k_+(y_i)a(2x_i)} \sum_{r\in R_L} \sgn(r) \prod_{i=1}^L k_+(x_i^r) \nonumber\\
\hphantom{\mathcal{Z}_\text{tri}(\vec{x};\vec{y}\,)=}{}\times \prod_{i<j}^L \frac{b(x_i^r+x_j^r)}{a(x_i^r+x_j^r)} \prod_{i,j=1}^L a(x_i^r + y_j) \times Z(\vec{x}\,{}^r;\vec{y}\,) ,\label{eq:new_formula_6v}
	\end{gather}
with $k_\pm(w) = \sin\gamma(\kappa\pm w)$ the six-vertex limit of \eqref{eq:vertices_dynamical_boundary} and $a$, $b$ as in \eqref{eq:vertices}. Since in the homogeneous limit $x_i \to x$ we get $a(x_i^r+x_j^r)\to 0$ whenever $r=r_I$ with $\#(I \cap \{i,j\})=1$ it is not clear whether \eqref{eq:new_formula_6v} might facilitate the computation of that limit. When we instead take the rational limit $[w]\to w$, where one may set $\gamma=1$, we recover the result of the boundary perimeter Bethe ansatz~\cite{Fra_17} for the lattice \eqref{eq:partition_function_SOS}.\footnote{See the case $m=L$ of~(80) in \cite{Fra_17}, with $z_i\coloneqq x_i$, $v_j\coloneqq {-y_j}$, $q\coloneqq \kappa$ and finally $\tau = \tau_I \coloneqq r_{\{1,\dots,L\}\setminus I}$, which is equivalent to always picking up the \emph{second} term from \eqref{eq:m_n} to rewrite \eqref{eq:symmetrized_sum_dyn}.}

\section{Summary and outlook}\label{s:concl}

The main purpose of this review is to advertise the results contained in the author's thesis~\cite{Lam_16}, and especially those that were not published before, on the functional method of Galleas and the author. However, we also presented some new results, see especially the end of Section~\ref{s:SOS}. In particular we have obtained the more compact expression \eqref{eq:new_formula_SOS}--\eqref{eq:new_formula_SOS_rewrite} for the partition function of the solid-on-solid model with domain walls and one reflecting end in terms of a~crossing-symmetrized sum with $2^L$ terms featuring the elliptic domain-wall partition function of \cite{PRS_08,Ros_09}. In the trigonometric case of the six-vertex model our result boils down to a relation between the reflecting-end partition function and the domain-wall partition function, see \eqref{eq:new_formula_6v}, which to the best of our knowledge is new, and can be matched with the outcome of the boundary perimeter Bethe ansatz by Frassek~\cite{Fra_17}.

Concerning the functional method itself, our main messages are that
\begin{itemize}\itemsep=0pt
	\item it contains the approach of Korepin--Izergin, cf.~\eqref{eq:functional_implies_KI}, as was shown in~\cite{Lam_16};
	\item it provides a recipe to get a direct formula for the partition function~\cite{Gal_11}: in some sense Lagrange interpolation is `built in';
	\item it can be made rigorous, as was done for the domain-wall partition function in~\cite{Lam_16};
	\item it is fairly general: it can for example also be applied to the elliptic solid-on-solid model with domain walls and one reflecting end~\cite{Lam_15,Lam_16}.
\end{itemize}
To stress the second point we used the terminology `constructive method' in \cite{Lam_16}. For the domain-wall case we have moreover compared the functional method to the solution of Korepin's recurrence relations by Lagrange interpolation and related approaches~\cite{BPZ_02}.

To date all cases in which the functional method has been used to obtained a closed expression were previously tackled using the Korepin--Izergin method. For the six-vertex model the domain-wall partition function~\cite{Gal_10,Lam_16} was of course first found by Korepin and Izergin \cite{Ize_87,ICK_92, Kor_82}, the reflecting-end partition function~\cite{GL_14} by Tsuchiya~\cite{Tsu_98}, and the corresponding (off-/on-shell) scalar products of Bethe vectors~\cite{Gal_14,Gal_15b} by Slavnov~\cite{Sla_89}, Wang~\cite{Wan_02} and Kitanine et al.~\cite{KK+_07}. In the dynamical (\textsc{sos}) case the domain-wall partition function~\cite{Gal_12,Gal_13a} was found by Rosengren~\cite{Ros_09}, and the reflecting-end partition function~\cite{Lam_15,Lam_16} by Filali and Kitanine~\cite{Fil_11b,Fil_11a, FK_10}.

Of course the real challenge of the functional method is to apply it to situations that have not been tackled before. Unfortunately the computation of the domain-wall partition function of the nineteen-vertex model~\cite{Gar_16} seems out of reach, cf.~the end of Section~\ref{s:applicability}. Another opportunity that comes to mind is the computation of $n$-point correlators, which can be interpreted as partition functions of lattices where the arrows on $n$ edges are fixed, cf.~\cite{BPZ_02}. Applications in a different direction, for which the functional equations themselves~-- rather than the resulting recipes for recursion on which we have focussed in this text~-- are crucial, are being developed by Galleas~\cite{Gal_16b,Gal_16a,Gal_16u,Gal_17,Gal_17u}.

Finally we cannot help noticing that the functional method provides a beautiful example of the rigid structure imposed by the underlying algebra, which is reflected in the many remarkable properties of the functional equations obtained in this way.

\subsection*{Acknowledgements}

This text has grown out of a presentation delivered at the \textit{Les Houches Summer School on Integrability} in June 2016 and a poster presentation at the \textit{ESI Workshop Elliptic Hypergeometric Functions in Combinatorics, Integrable Systems and Physics} in March~2017. It is a pleasure to thank the organizers of these excellent meetings. I am grateful to W.~Galleas for collaboration on~\cite{GL_14} and for suggesting the problem addressed in~\cite{Lam_15}. I thank G.~Arutyunov and A.~Henriques for discussions while working on~\cite{Lam_15}, H.~Rosengren for detailed feedback on~\cite{Lam_16} and the present text, and A.~Garbali and R.A.~Pimenta for trying the functional method for the nineteen-vertex model.

For the revised version of this review I further thank the anonymous referee for valuable feedback, A.G.~Pronko for correspondence, and H.~Rosengren for various useful discussions.

The works \cite{GL_14,Lam_15,Lam_16} were supported by the \textsc{vici} grant 680-47-602 and by the \textsc{erc} Advanced Grant 246974, \textit{Supersymmetry: a window to non-perturbative physics}, and part of the \textsc{d-itp} consortium, an \textsc{nwo} program funded by the Dutch Ministry of Education, Culture and Science (\textsc{ocw}). The present review was written with the support from the Knut and Alice Wallenberg Foundation (\textsc{kaw}).

\pdfbookmark[1]{References}{ref}
\LastPageEnding


\begin{thebibliography}{99}
\footnotesize\itemsep=0pt

\bibitem{Bax_87}
Baxter R.J., Perimeter {B}ethe ansatz, \href{https://doi.org/10.1088/0305-4470/20/9/039}{\textit{J.~Phys.~A: Math. Gen.}}
 \textbf{20} (1987), 2557--2567.

\bibitem{BPZ_02}
Bogoliubov N.M., Pronko A.G., Zvonarev M.B., Boundary correlation functions of
 the six-vertex model, \href{https://doi.org/10.1088/0305-4470/35/27/301}{\textit{J.~Phys.~A: Math. Gen.}} \textbf{35} (2002),
 5525--5541, \href{https://arxiv.org/abs/math-ph/0203025}{math-ph/0203025}.

\bibitem{Che_84}
Cherednik I.V., Factorizing particles on a half line, and root systems,
 \href{https://doi.org/10.1007/BF01038545}{\textit{Theoret. and Math. Phys.}} \textbf{61} (1984), 977--983.

\bibitem{Fil_11b}
Filali G., Dynamical reflection algebra and associated boundary integrable
 models, Ph.D.~Thesis, Universit\'e de Cergy Pontoise, 2011, available at
 \url{https://tel.archives-ouvertes.fr/tel-00664076}.

\bibitem{Fil_11a}
Filali G., Elliptic dynamical reflection algebra and partition function of
 {SOS} model with reflecting end, \href{https://doi.org/10.1016/j.geomphys.2011.01.002}{\textit{J.~Geom. Phys.}} \textbf{61} (2011),
 1789--1796, \href{https://arxiv.org/abs/1012.0516}{arXiv:1012.0516}.

\bibitem{FK_10}
Filali G., Kitanine N., The partition function of the trigonometric {SOS} model
 with a reflecting end, \href{https://doi.org/10.1088/1742-5468/2010/06/L06001}{\textit{J.~Stat. Mech. Theory Exp.}} \textbf{2010}
 (2010), L06001, 11~pages, {E}rratum,
 \href{https://doi.org/10.1088/1742-5468/2010/07/E07002}{\textit{J.~Stat.
 Mech. Theory Exp.}} \textbf{2010} (2010), E07002, \href{https://arxiv.org/abs/1004.1015}{arXiv:1004.1015}.

\bibitem{Fra_17}
Frassek R., Boundary perimeter {B}ethe ansatz, \href{https://doi.org/10.1088/1751-8121/aa7278}{\textit{J.~Phys.~A: Math.
 Theor.}} \textbf{50} (2017), 265202, 19~pages, \href{https://arxiv.org/abs/1703.10842}{arXiv:1703.10842}.

\bibitem{Gal_10}
Galleas W., Functional relations for the six-vertex model with domain wall
 boundary conditions, \href{https://doi.org/10.1088/1742-5468/2010/06/P06008}{\textit{J.~Stat. Mech. Theory Exp.}} \textbf{2010} (2010), P06008,
 15~pages, \href{https://arxiv.org/abs/1002.1623}{arXiv:1002.1623}.

\bibitem{Gal_11}
Galleas W., A new representation for the partition function of the six-vertex
 model with domain wall boundaries, \href{https://doi.org/10.1088/1742-5468/2011/01/P01013}{\textit{J.~Stat. Mech. Theory Exp.}}
\textbf{2011} (2011), P01013, 12~pages, \href{https://arxiv.org/abs/1010.5059}{arXiv:1010.5059}.

\bibitem{Gal_12}
Galleas W., Multiple integral representation for the trigonometric {SOS} model
 with domain wall boundaries, \href{https://doi.org/10.1016/j.nuclphysb.2012.01.006}{\textit{Nuclear Phys.~B}} \textbf{858} (2012),
 117--141, \href{https://arxiv.org/abs/1111.6683}{arXiv:1111.6683}.

\bibitem{Gal_13b}
Galleas W., Functional relations and the {Y}ang--{B}axter algebra,
 \href{https://doi.org/10.1088/1742-6596/474/1/012020}{\textit{J.~Phys. Conf. Ser.}} \textbf{474} (2013), 012020, 19~pages,
 \href{https://arxiv.org/abs/1312.6816}{arXiv:1312.6816}.

\bibitem{Gal_13a}
Galleas W., Refined functional relations for the elliptic {SOS} model,
 \href{https://doi.org/10.1016/j.nuclphysb.2012.10.014}{\textit{Nuclear Phys.~B}} \textbf{867} (2013), 855--871, \href{https://arxiv.org/abs/1207.5283}{arXiv:1207.5283}.

\bibitem{Gal_14}
Galleas W., Scalar product of {B}ethe vectors from functional equations,
 \href{https://doi.org/10.1007/s00220-014-1976-2}{\textit{Comm. Math. Phys.}} \textbf{329} (2014), 141--167, \href{https://arxiv.org/abs/1211.7342}{arXiv:1211.7342}.

\bibitem{Gal_15b}
Galleas W., Off-shell scalar products for the {$XXZ$} spin chain with open
 boundaries, \href{https://doi.org/10.1016/j.nuclphysb.2015.02.010}{\textit{Nuclear Phys.~B}} \textbf{893} (2015), 346--375,
 \href{https://arxiv.org/abs/1412.5389}{arXiv:1412.5389}.

\bibitem{Gal_16b}
Galleas W., Elliptic solid-on-solid model's partition function as a single
 determinant, \href{https://doi.org/10.1103/PhysRevE.94.010102}{\textit{Phys. Rev.~E}} \textbf{94} (2016), 010102, 5~pages,
 \href{https://arxiv.org/abs/1604.01223}{arXiv:1604.01223}.

\bibitem{Gal_16a}
Galleas W., New differential equations in the six-vertex model,
 \href{https://doi.org/10.1088/1742-5468/2016/03/033106}{\textit{J.~Stat. Mech. Theory Exp.}} \textbf{2016} (2016), 033106, 13~pages,
 \href{https://arxiv.org/abs/1508.04690}{arXiv:1508.04690}.

\bibitem{Gal_16u}
Galleas W., On the elliptic $\mathfrak{gl}_2$ solid-on-solid model: functional
 relations and determinants, \href{https://arxiv.org/abs/1606.06144}{arXiv:1606.06144}.

\bibitem{Gal_17}
Galleas W., Continuous representations of scalar products of {B}ethe vectors,
 \href{https://doi.org/10.1063/1.4997156}{\textit{J.~Math. Phys.}} \textbf{58} (2017), 083504, 15~pages,
 \href{https://arxiv.org/abs/1607.08524}{arXiv:1607.08524}.

\bibitem{Gal_17u}
Galleas W., Six-vertex model and non-linear differential equations I.~Spectral
 problem, \href{https://arxiv.org/abs/1705.03408}{arXiv:1705.03408}.

\bibitem{GL_14}
Galleas W., Lamers J., Reflection algebra and functional equations,
 \href{https://doi.org/10.1016/j.nuclphysb.2014.07.016}{\textit{Nuclear Phys.~B}} \textbf{886} (2014), 1003--1028, \href{https://arxiv.org/abs/1405.4281}{arXiv:1405.4281}.

\bibitem{Gar_16}
Garbali A., The domain wall partition function for the {I}zergin--{K}orepin
 nineteen-vertex model at a root of unity, \href{https://doi.org/10.1088/1742-5468/2016/03/033112}{\textit{J.~Stat. Mech. Theory Exp.}}
\textbf{2016} (2016), 033112, 19~pages, \href{https://arxiv.org/abs/1411.2903}{arXiv:1411.2903}.

\bibitem{GK_00}
G\"ohmann F., Korepin V.E., Solution of the quantum inverse problem,
 \href{https://doi.org/10.1088/0305-4470/33/6/308}{\textit{J.~Phys.~A: Math. Gen.}} \textbf{33} (2000), 1199--1220.

\bibitem{Ize_87}
Izergin A.G., Partition function of a six-vertex model in a finite volume,
 \textit{Sov. Phys. Dokl.} \textbf{32} (1987), 878--879.

\bibitem{ICK_92}
Izergin A.G., Coker D.A., Korepin V.E., Determinant formula for the six-vertex
 model, \href{https://doi.org/10.1088/0305-4470/25/16/010}{\textit{J.~Phys.~A: Math. Gen.}} \textbf{25} (1992), 4315--4334.

\bibitem{KK+_07}
Kitanine N., Kozlowski K.K., Maillet J.M., Niccoli G., Slavnov N.A., Terras V.,
 Correlation functions of the open {$XXZ$} chain.~{I}, \href{https://doi.org/10.1088/1742-5468/2007/10/P10009}{\textit{J. Stat. Mech.
 Theory Exp.}} \textbf{2007} (2007), P10009, 37~pages, \href{https://arxiv.org/abs/0707.1995}{arXiv:0707.1995}.

\bibitem{KMT_99}
Kitanine N., Maillet J.M., Terras V., Form factors of the {$XXZ$} {H}eisenberg
 spin-{$\frac 12$} finite chain, \href{https://doi.org/10.1016/S0550-3213(99)00295-3}{\textit{Nuclear Phys.~B}} \textbf{554} (1999),
 647--678, \href{https://arxiv.org/abs/math-ph/9807020}{math-ph/9807020}.

\bibitem{KMT_00}
Kitanine N., Maillet J.M., Terras V., Correlation functions of the {$XXZ$}
 {H}eisenberg spin-{${1\over2}$} chain in a~magnetic field, \href{https://doi.org/10.1016/S0550-3213(99)00619-7}{\textit{Nuclear
 Phys.~B}} \textbf{567} (2000), 554--582, \href{https://arxiv.org/abs/math-ph/9907019}{math-ph/9907019}.

\bibitem{Kor_82}
Korepin V., Calculation of norms of {B}ethe wave functions, \href{https://doi.org/10.1007/BF01212176}{\textit{Comm. Math.
 Phys.}} \textbf{86} (1982), 391--418.

\bibitem{KZ_00}
Korepin V., Zinn-Justin P., Thermodynamic limit of the six-vertex model with
 domain wall boundary conditions, \href{https://doi.org/10.1088/0305-4470/33/40/304}{\textit{J.~Phys.~A: Math. Gen.}} \textbf{33}
 (2000), 7053--7066, \href{https://arxiv.org/abs/cond-mat/0004250}{cond-mat/0004250}.

\bibitem{Lam_14}
Lamers J., A pedagogical introduction to quantum integrability, with a view
 towards theoretical high-energy physics, \textit{PoS Proc. Sci.} (2014),
 PoS(Modave2014), 001, 70~pages, \href{https://arxiv.org/abs/1501.06805}{arXiv:1501.06805}.

\bibitem{Lam_15}
Lamers J., Integral formula for elliptic {SOS} models with domain walls and a
 reflecting end, \href{https://doi.org/10.1016/j.nuclphysb.2015.11.006}{\textit{Nuclear Phys.~B}} \textbf{901} (2015), 556--583,
 \href{https://arxiv.org/abs/1510.00342}{arXiv:1510.00342}.

\bibitem{Lam_16}
Lamers J., On elliptic quantum Integrability: vertex models, solid-on-solid
 models and spin chains, Ph.D.~Thesis, Utrecht University, 2016, available at
 \url{https://dspace.library.uu.nl/handle/1874/333998}.

\bibitem{Lie_67b}
Lieb E.H., Exact solution of the {$F$} model of an antiferroelectric,
 \href{https://doi.org/10.1103/PhysRevLett.18.1046}{\textit{Phys. Rev. Lett.}} \textbf{18} (1967), 1046--1048.

\bibitem{Lie_67c}
Lieb E.H., Exact solution of the two-dimensional {S}later {KDP} model of a
 ferroelectric, \href{https://doi.org/10.1103/PhysRevLett.19.108}{\textit{Phys. Rev. Lett.}} \textbf{19} (1967), 108--110.

\bibitem{Lie_67a}
Lieb E.H., Residual entropy of square ice, \href{https://doi.org/10.1103/PhysRev.162.162}{\textit{Phys. Rev.}} \textbf{162}
 (1967), 162--172.

\bibitem{MT_00}
Maillet J.M., Terras V., On the quantum inverse scattering problem,
 \href{https://doi.org/10.1016/S0550-3213(00)00097-3}{\textit{Nuclear Phys.~B}} \textbf{575} (2000), 627--644,
 \href{https://arxiv.org/abs/hep-th/9911030}{hep-th/9911030}.

\bibitem{PRS_08}
Pakuliak S., Rubtsov V., Silantyev A., The {SOS} model partition function and
 the elliptic weight functions, \href{https://doi.org/10.1088/1751-8113/41/29/295204}{\textit{J.~Phys.~A: Math. Theor.}} \textbf{41}
 (2008), 295204, 20~pages, \href{https://arxiv.org/abs/0802.0195}{arXiv:0802.0195}.

\bibitem{Ros_09}
Rosengren H., An {I}zergin--{K}orepin-type identity for the 8{VSOS} model, with
 applications to alternating sign matrices, \href{https://doi.org/10.1016/j.aam.2009.01.003}{\textit{Adv. in Appl. Math.}}
 \textbf{43} (2009), 137--155, \href{https://arxiv.org/abs/0801.1229}{arXiv:0801.1229}.

\bibitem{Skl_88}
Sklyanin E.K., Boundary conditions for integrable quantum systems,
 \href{https://doi.org/10.1088/0305-4470/21/10/015}{\textit{J.~Phys.~A: Math. Gen.}} \textbf{21} (1988), 2375--2389.

\bibitem{Sla_89}
Slavnov N.A., Calculation of scalar products of wave functions and form-factors
 in the framework of the algebraic {B}ethe ansatz, \href{https://doi.org/10.1007/BF01016531}{\textit{Theoret. and Math.
 Phys.}} \textbf{79} (1989), 502--508.

\bibitem{Sut_67}
Sutherland B., Exact solution of a two-dimensional model for hydrogen-bonded
 crystals, \href{https://doi.org/10.1103/PhysRevLett.19.103}{\textit{Phys. Rev. Lett.}} \textbf{19} (1967), 103--104.

\bibitem{TV_97}
Tarasov V., Varchenko A., Geometry of {$q$}-hypergeometric functions, quantum
 affine algebras and elliptic quantum groups, \textit{Ast\'erisque}
 \textbf{246} (1997), vi+135~pages, \href{https://arxiv.org/abs/q-alg/9703044}{q-alg/9703044}.

\bibitem{Tsu_98}
Tsuchiya O., Determinant formula for the six-vertex model with reflecting end,
 \href{https://doi.org/10.1063/1.532606}{\textit{J.~Math. Phys.}} \textbf{39} (1998), 5946--5951,
 \href{https://arxiv.org/abs/solv-int/9804010}{solv-int/9804010}.

\bibitem{Wan_02}
Wang Y.S., The scalar products and the norm of {B}ethe eigenstates for the
 boundary {$XXX$} {H}eisenberg spin-1/2 finite chain, \href{https://doi.org/10.1016/S0550-3213(01)00610-1}{\textit{Nuclear Phys.~B}}
 \textbf{622} (2002), 633--649.

\bibitem{Zin_00}
Zinn-Justin P., Six-vertex model with domain wall boundary conditions and
 one-matrix model, \href{https://doi.org/10.1103/PhysRevE.62.3411}{\textit{Phys. Rev.~E}} \textbf{62} (2000), 3411--3418,
 \href{https://arxiv.org/abs/math-ph/0005008}{math-ph/0005008}.

\end{thebibliography}
\end{document}